\documentclass[twocolumn]{aastex631}
\usepackage[utf8]{inputenc}
\usepackage{mathtools}
\usepackage{comment}
\published{August 2023}
\submitjournal{ApJ}

\shorttitle{Stellar Collisions in the GC}
\shortauthors{Balberg \& Yassur}

\newcommand{\pfrac}[2]{\left( \frac{#1}{#2} \right)}

\newcommand{\MBH}{M_{\bullet}}
\newcommand{\Msun}{M_{\odot}}

\newcommand{\e}{\varepsilon}

\newcommand{\ttb}{T_{\rm 2B}}
\newcommand{\ttbj}{T^J_{\rm 2B}}

\newcommand{\invyr}{yr^{\rm {-1}}}

\submitjournal{\apj}

\begin{document}
\title{Stellar Collisions in Galactic Nuclei: Impact on Destructive Events Near a Supermassive Black Hole}

\correspondingauthor{Shmuel Balberg}
\email{shmuel.balberg@mail.huji.ac.il}

\author[0000-0002-0349-1875]{Shmuel Balberg}
\affiliation{Racah Institute, Hebrew University of Jerusalem, Jerusalem, 91904, Israel}

\author{Gilad Yassur}
\affiliation{Racah Institute, Hebrew University of Jerusalem, Jerusalem, 91904, Israel}

\begin{abstract}
The centers of galaxies host both a supermassive black hole and a dense stellar cluster. Such an environment should lead to stellar collisions, possibly at very high velocities so that the total energy involved is of the same order as supernova explosions. We present a simplified numerical analysis of the destructive stellar collision rate in a cluster similar to that of the Milky Way. The analysis includes an effective average two-body relaxation Monte Carlo scheme and general relativistic effects, as used by \citet{SariFragione2019}, to which we added explicit tracking of local probabilities for stellar collisions. We also consider stars which are injected into the stellar cluster after being disrupted from a binary system by the supermassive black hole. Such stars are captured in the vicinity of the black hole and enhance the expected collision rate. In our results we examine the rate and energetic distribution function of high-velocity stellar collisions, and compare them self-consistently with the other destructive processes which occur in the galactic center, namely tidal disruptions and extreme mass ratio inspirals.
\end{abstract}


\section{Introduction\label{sec:Intro}}

The presence of supermassive black holes in the centers of galaxies appears to be ubiquitous in the universe \citep{KormendyHo2013}. Such black holes regulate galaxy formation \citep{HeckmanBest2014}, and drive radiative emission in active galactic nuclei \citep{AGNI,AGNII}. A central black hole also determines many of the properties of the dense stellar cluster that surrounds it near the galactic center. The structure and dynamics of the cluster are a combined result of the black hole dominating the potential, along with a multitude of gravitational interactions amongst the stars \citep{RauchTremaine1996,HopmanAlexander2005,Merritt2013,Alexander2017,Fouvryetal2022}. This combination tends to generate a particular spatial density profile in the stellar cluster, which evolves dynamically through gravitational relaxation processes. 

A unique aspect of the dense stellar cluster is the destructive events which its stars experience. Much attention has been devoted to tidal disruption events (TDEs), when a star approaches the supermassive black hole (SMBH) at a distance $R_T$, where the tidal force of the black hole overpowers the gravitational binding of the star. Such an event is possible when $R_T$ lies outside of the event horizon, and possibly leads to a highly luminous transient, as the stellar matter interacts with itself while accreting onto the black hole \citep{Rees1988,Gezari2021}. 
 
As noted before by several authors (see, e.g. \citet{Alexander2017} for a review), there are actually two channels that lead a star to tidal disruption. First, stars throughout the cluster can be gravitationally scattered by other stars into highly eccentric orbits, whose periapse lies below $R_T$. Such scatterings can occur at any distance from the SMBH, including stars at the radius of influence, $R_h$, which are usually the most abundant. Hereafter we will refer to this channel as a TDE. The second channel is due to general relativistic effects, in which gravitational wave (GW) losses cause single stars to inspiral gradually toward the SMBH. Initially GW losses tend to circularize the orbit while keeping an almost fixed periapse, followed by a gradual inspiral toward $R_T$ along roughly circular orbits. While technically this inspiral also terminates with the star being tidally disrupted, the process occurs over multiple orbits causing a more "gentle" disruption. This latter channel is referred to as a main-sequence extreme mass ratio inspiral (EMRI), and is mostly studied in the context of potential sources of observable GW emission (along with EMRIs of compact objects which cannot be tidally disrupted and therefore spiral all the way to the event horizon). However, given that a significant fraction of the star's mass may shed prior to disruption, optical transients are also possible \citep{LinialSari2023}, especially if more than one star can undergo an EMRI simultaneously \citep{MetzgerStone2017,Metzgeretal2022}.

The rates of TDEs and EMRIs  depend on the properties of the SMBH and the stellar cluster. In both cases, the rate is determined by gravitational scattering among the stars, which repopulate orbits that terminate in disruption. In general, the typical estimated TDE rate for a Milky Way-like SMBH and star cluster is of order $10^{-5} {\invyr}$. The EMRI rate (of main-sequence stars) in a fully relaxed cluster with Milky Way properties is estimated to be about $1\%$ of the TDE rate, but as shown by \cite{SariFragione2019}, this rate is significantly enhanced when disrupted binaries are taken into account (see below).

Relatively little attention has been given to destructive stellar collisions, which are also possible in a dense cluster surrounding an SMBH. Physical collisions in the dense stellar cluster should actually be common, but mostly at large distances from the black hole, with velocities which are smaller than the stars' escape velocities. Such collisions are more likely to result in one or both stars surviving or merging \citep{FreitagBenz2005}. Mass loss, mass transfer, and mergers in such collisions should reshape the mass function of the stellar population \citep{Roseetal2023}. In particular, collisions are suggested as a possible formation mechanism for young, high-mass stars, such as S-stars observed surrounding Sagittarius A \citep{FragioneAntonini2018}. However, closer to the SMBH, where the orbital velocities reach thousands of kilometers per second, collisions are completely destructive, and could led to a variety of potential observational consequences \citep{AmaroSeoane2023}. 

A fast, head-on collision of this sort will result in hot expanding gas, implying that it should produce some form of an optical flare. In fact, at the higher end of the velocity range, collisions between sun-like stars at 10,000 km S$^{-1}$ contain kinetic energies of $10^{51}$ erg or more, which conform with supernova-like energy output. Accordingly, \cite{BalbergSariLoeb2013} suggested that such hypervelocity collisions should be considered as another, rare, form of supernova. A rough estimate suggests that the resulting optical flare should indeed be "supernova-like" over a time scale of order a few days (and decay quickly given that no radioactive isotopes are produced). A second luminous transient might be possible if there is significant fast accretion of the debris onto the SMBH  

Clearly destructive collisions (DCs) are inevitable, but at what rate? How do they compare, and possibly compete, with TDEs and EMRIs as stellar destruction scenarios near the SMBH? The answers depend on the steady-state distribution of stars in the cluster surrounding the SMBH. In turn, this distribution strongly depends on the mechanisms that supply stars to orbits which are relevant for destruction in the different scenarios. 

In this work, we present a simplified numerical study of the expected rates of DCs in a "Milky Way-like" galactic center. We approximate the stellar cluster as being composed of identical mass $m$ stars orbiting the SMBH, and track the evolution of the cluster with a simplified Monte Carlo simulation. This scheme was originally presented by \cite{SariFragione2019} for TDEs and EMRIs, to which we added DCs. In the simulations we include stars which originated from disrupted tight binaries by the SMBH \citep{Hills1988}, and captured in highly eccentric orbits with relatively small ($1-100R_T$) periapses. The existence of such captured stars is inferred from observations of hypervelocity stars ejected from the Milky Way \citep{Brown2015}, which are believed to be companions that became unbound during the binary disruption. As shown by \cite{BalbergSariLoeb2013}, these captured stars are prime candidates for DCs, and in fact could make their rate comparable to that of TDEs. Stars from disrupted binaries obviously also affect the TDE and EMRI rates \citep{SariFragione2019}, and the main goal of the present work is to study all three in a self-consistent calculation. 

The structure of this manuscript is as follows. In Section \ref{sec:Basics}, we review the physics of two-body relaxation, GW emission, and collisions and evaluate their time scales in a schematic (power-law density) cluster. The equations and setup of our Monte Carlo simulation, following \cite{SariFragione2019} and the introduction of the possibility for collisions, are presented in Section \ref{sec:MCpluscollisions}. In Section \ref{sec:Resultsnobinaries} we examine the results of the simulations, and most notably the rates of the different destruction scenarios when no stars from disrupted binaries are included. Such stars are added to the simulations in Section \ref{sec:Resultswithbinaries}, where we demonstrate their dramatic enhancement of the DC rate. Finally, in Section \ref{sec:Summary} , we discuss the implications of the results and our corresponding conclusions.

\section{Principal processes in the dense stellar cusp} \label{sec:Basics}

A stellar cluster surrounding an SMBH is expected to form a dense cusp, with the density monotonically increasing toward the center. The conventional approach is to identify the SMBH radius of influence, $R_h$, at which the total enclosed mass of the stars is equal to the SMBH mass, $\MBH$. Inside of $R_h$ we approximate that the SMBH dominates the gravitational potential, and that at every instant the motion of each individual star is consistent with a Keplerian orbit in this potential. In this work we set all the stars in the cusp to have identical mass $m$, so the number of stars enclosed by the radius of influence is simply $N(\leq R_h)=\MBH/m$.

The orbits of the stars do evolve over time with respect to a stationary Keplerian orbit. The two main driving processes are gravitational scattering among the stars and GW emission. Both processes can be estimated by approximating the stars as point masses. 

Stellar collisions introduce an additional mechanism which can change both the mass and the number of stars. Low-velocity collisions (and even near misses) far from the SMBH can result in mass transfer, mass loss, and mergers, while  hypervelocity collisions at small distances from the SMBH remove stars through complete destruction in a singe event. The prospects and physics of collisions depend on the sizes of the stars, which we treat here as identical with radius $R_\star$. 

\subsection{Gravitational scattering and two-body relaxation
\label{subsec:twobodyNR}}

While the SMBH dominates the gravitational potential, stars also interact with each other. Any given star experiences mostly weak, small-angle scatterings with other stars, but rare strong scatterings at small impact parameters occur as well, and have a similar effect on the total accumulated changes in the specific energy, $E$, and the specific angular momentum, $J$ \citep{LightmanShapiro1977}. Under the simplest assumption that all scatterings are random, the star's orbital parameters change diffusively in phase space. A star orbiting the SMBH with a semi-major-axis (sma) $r$ will change its specific angular momentum by order unity over a two-body relaxation time, given by \citep{BW76,BW77}

\begin{equation}\label{eq:T2B}
    \ttb(r) \approx \frac{P(r)}{N(r) \ln{\Lambda}} \pfrac{\MBH}{m}^2\,.
\end{equation}

Here $P(r)$ is the orbital period of a Keplerian orbit around the SMBH, $N(r)$ is the number of stars in the proximity of the radius $r$ and $\ln{\Lambda}$ is the resulting Coulomb logarithm. Assuming that most stars have roughly circular orbits, $N(r)$ is approximately the number of stars enclosed by radius $r$. The time scale for order-of-unity diffusion in energy space is somewhat longer than $\ttb(r)$, so stars generally change their periapse prior to changing their smas significantly. 

This two-body relaxation time allows us to derive a rough estimate for the TDE rate \citep{SariFragione2019}. Assuming that the cluster is dominated by stars with an sma $r\sim R_h$ and that at $R_h$ the velocity distribution is isotropic, TDEs reflect the rate at which stars at the radius of influence are scattered into orbits with angular momentum $J\leq J_{LC}=\sqrt{2G\MBH R_T}$.  A system with Milky Way-like parameters will tend to an "empty-loss-cone" scenario, in which there is on average less than one star in an orbit that terminates in tidal disruption. Hence the TDE rate is set by stars evolving from a circular orbit to a highly eccentric one that will "plunge" below $R_T$. This occurs as a diffusive process in two-dimensional angular momentum space, originating from the circular angular momentum $J_c(R_h)=\sqrt{G\MBH R_h}$ and terminating at the loss-cone value $J_{LC}$. The TDE rate should be of order
\begin{equation}\label{eq:rateTDEsimp}
    \mathcal{R}_{TDE}\approx 
    \frac{N(R_h)}{\ln(J_C(R_h)/J_{LC})\ttb(R_h)}
    \approx
     \frac{1}{P(R_h)}\,,
\end{equation}
since by construction $N(R_h)=\MBH/m$, and we assume that $\ln(J_c(R_h)/J_{LC})$ and $\ln\Lambda$ have similar values (of order 10). For the Milky Way  $P(R_h)=2\pi(R^3_h/G\MBH)^{1/2}\approx1.3\times10^5\;$yr, implying a TDE rate of about $7.6\times 10^{-6}\;\invyr$. 

For our purpose of studying the steady-state rates of all destructive events which occur close to the SMBH, we also consider the evolution of orbits. A star on an eccentric orbit with sma $r$ and periapse distance $r_p$ will change its specific angular momentum by order unity over a time scale smaller than $\ttb(r)$ by a factor of $r_p/r\approx(J/J_c(r))^2$. 
The eccentric orbital relaxation time is therefore
\citep{BinneyTremaine1987}
\begin{equation}\label{eq:T2BJ}
    \ttbj(r,r_p) \approx \frac{P(r)}{N(r) \ln{\Lambda}} \pfrac{\MBH}{m}^2\left(\frac{r_p}{r}\right)\,,
\end{equation}
which is appropriate if the density profile is not too steep, so that stars at $r$ dominate the scatterings (this assumption applies as long as the density of the stars, $n(r)$, is shallower than $n(r)\sim r^{-4}$, which we shall assume henceforth).

In this work we follow the standard low-order approximation of gravitational dynamics between stars, and limit ourselves to random two-body scatterings. In reality the dynamics between stars are subject to other relaxation mechanisms, commonly referred to as resonant relaxation \citep{RauchTremaine1996,Alexander2017}. These arise from residual torques between the stars, which lead to nondiffusive evolution of the angular momenta, and over time scales which are significantly shorter than $\ttb$ \citep{KocsisTremaine2011,KocsisTremaine2015}. On the other hand, the effect of resonant relaxation is restricted by the timescales over which a star experiences coherent fluctuations in the gravitational potential \citep{HopmanAlexander2006}. Detailed analysis \citep{BarOrAlexander2016} shows that random two-body scattering are sufficient to suppress coherent effects far from the SMBH, and the same occurs close to the SMBH due to general relativistic effects. As a result, resonant relaxation in a Milky Way-like cusp is limited to intermediate distances from the SMBH. Since TDEs are mostly the result of stars being scattered into eccentric orbits close to $R_h$, and EMRIs arise from general relativistic effects close to the SMBH, resonant relaxation was found to have little impact on their rates. Our focus is on collisions in regions close to the SMBH, so we expect that neglecting resonant relaxation is a legitimate approximation for our analysis as well. 

\subsection{Orbital evolution through GW emission\label{subsec:GW}}

Gravitational wave emission by the stars in the background field of the SMBH is a dissipation mechanism. It drives an orbit toward a smaller sma, along with a slower decay of the periapse distance \citep{Peters1964}. As a result, GW emission will tend to circularize eccentric orbits, and then drive a gradual inspiral of roughly circular orbits to the innermost stable relativistic orbit (ISCO). The time scale for a full inspiral for an orbit initially with sma $r$ and periapse distance $r_p$ is \citep{Peters1964,HopmanAlexander2005}
\begin{equation}
T_{GW}(r,r_p)=\frac{R_S}{c}\frac{\MBH}{m} \left(\frac{r_p}{R_S} \right)^4 \left(\frac{r}{r_p} \right)^{1/2}\ ,
\label{eqn:gw}
\end{equation}
where $R_S=2G\MBH/c^2$ is the Schwarzchild radius. This result applies when $T_{GW}(r,r_p)$ is longer than the star's orbital period, $P(r)$. This condition is easily satisfied for any $r<R_h$ \citep{SariFragione2019}, and while it is only marginally satisfied at $r\approx R_h$, the general relativistic effects there are small enough to be ignored in relevant calculations. 

As noted by \cite{SariFragione2019}, there exists a region in the $r-r_p$ plane for which $T_{GW}(r,r_p)<\ttbj(r,r_p)$, depending, of course, on the steady-state stellar density profile, $N(r)$. Orbits in this region will evolve toward an EMRI, rather than toward a TDE. Specifically, if the steady state follows the \cite{BW76} power law of $N(r)\sim r^{5/4}$ (hereafter a "BW profile"), the $r_p(r)$ relation for which $T_{GW}(r,r_p)=\ttbj(r,r_p)$ has the form
\begin{equation}
\left. \frac{r_p}{R_S} \right. =(\ln \Lambda)^{-2/5}\left(\frac{r}{R_h}\right)^{-1/2}\ .
\label{eq:T2BJeqTGW}
\end{equation}
For stars with sma $r$ and a periapse $r_p$ smaller than that of equation \ref{eq:T2BJeqTGW}, the orbit will evolve into a series of quasi-circular orbits, and terminate as an EMRI.

This analysis yields an estimate of the EMRI rate \citep{SariFragione2019}. The largest influx of stars into combinations of $(r,r_p)$ at which GW dominates the orbital evolution is from eccentric orbits with an sma $r_0$ which satisfies equation \ref{eq:T2BJeqTGW} with $r_p=R_T$. This sma is $r_0=R_h(R_S/R_T)^2$, and in a similar fashion to equation \ref{eq:rateTDEsimp} (again, assuming a BW profile of $N(r)\sim r^{5/4}$)
\begin{equation}\label{eq:rateEMRIsimp}
\mathcal{R}_{EMRI}\approx\frac{N(r_0)}{log\left(\frac{J_C(r_0)}{J_{LC}(r_0)}\right)\ttb(r_0)}=\frac{1}{P(R_h)} \left(\frac{R_S}{R_T} \right)^{2}\ .
\end{equation}

For the Milky Way SMBH and sun-like stars, and assuming a BW density profile this results in $\mathcal{R}_{EMRI} \approx 10^{-2} \mathcal{R}_{TDE}$.

\subsection{Destructive stellar collisions\label{subsec:collisions}}

Stars will physically collide if their centers come within a distance $f_R R_\star$ of each other. Henceforth we track collisions only in the inner part of the cluster, where the orbital velocities are greater than the escape velocity of the stars. Correspondingly, we ignore gravitational focusing, so that $f_R\leq 2$. In the following we consider $f_R=0$ (no collisions), $f_R=1$ and $f_R=2$. The latter is an upper limit and probably an overestimate, since grazing collisions, even at hypervelocities, may not be completely disruptive. 

Stars on eccentric orbits sample different stellar densities along their path. In density profiles steeper than $n(r)\sim r^{-1}$ the optical depth for collisions is dominated by stars near the periapse  \citep{SariFragione2019} and the typical collision time for a star with orbital parameters $(r,r_p)$ is of order 
\begin{equation}
T_{\rm col}(r,r_p)=\frac{r_p^2 r}{N(r_p)} (f_R R_\star)^{-2} P(r)\;.
\label{eq:tcoll}
\end{equation}

Although they did not include collisions in their work, \cite{SariFragione2019} did point out that the collisional time scale (equation \ref{eq:tcoll}) is usually shorter than both the two-body and the GW timescales (for Milky Way parameters and assuming a BW cusp). Quantitatively, they assessed that collisions dominate the evolution for practically all orbits with $r<10^{-2}R_h$. 

Furthermore, consider the rate of hypervelocity collisions in a power-law density profile, $n(r)=A r^{-\alpha}$. Integrating over all radii in a range $r_{min}$ to $r_{max}$, the DC rate is
\begin{equation}\label{eq:rateCollpartI}
\mathcal{R}_{DC}=
\frac{1}{2}\int^{r_{max}}_{r_{min}}4\pi r^2 n^2(r) v(r) \pi (f_R R_\star)^2 dr \ ,
\end{equation}
where $v(r)\approx(G\MBH/r)^{1/2}$ is the typical velocity at radius $r$.
The constant $A$ is determined by the total amount of stars in the cluster $N\equiv N(R_h)$, $N=4\pi A R^{3-\alpha}_h/(3-\alpha)$. Correspondingly
\begin{equation}\label{eq:rateCollpartII}
\begin{split}
\mathcal{R}_{DC}=\frac{(3-\alpha)^2}{2(2\alpha-5/2)}\left(\frac{f_R}{2}\right)^2\frac{N^2(R_h)}{P(R_h)}\\
\times\left(\frac{R_\star}{R_h}\right)^2\left(\frac{R_{min}}{R_h}\right)^{5/2-2\alpha}\;
\end{split}
\end{equation}
(see also \citet{AmaroSeoane2023}).

Equation \ref{eq:rateCollpartII} applies for $5/4<\alpha<3$ so that $N$ is dominated by stars in the vicinity of $R_h$, but the collisions are dominated by the vicinity of $r_{min}$. For a BW profile with $\alpha=7/4$, Milky Way and sun-like values, and setting $r_{min}=R_T$, we find a dramatic rate of $\mathcal{R}_{DC}\approx 10^{-3}\;\invyr$ (for $f_R=1$). This result is unphysical and clearly indicates that collisions between stars will tend to deplete the inner part of the cusp. Depletion will dominate out to a radius $R_{col}$, where the orbital velocity is approximately equal to the escape velocity of the stars $v(r)\approx G\MBH/R_{col}=v_{esc}$. For the Milky Way and sun-like parameters this is roughly also the radius where $\ttb=T_{col}$ \citep{SariFragione2019}, so we expect that that the entire density profile will include a fully relaxed cusp (with a BW profile) which extends from $R_h$ into $R_{col}$, where DCs take over and the number of stars plummets. The collision rate is therefore determined by the BW density at $R_{col}$, $n(R_{col})\approx N(R_{col})/V(R_{col})$, where $V(R_{col})$ is the volume enclosed by $R_{col}$:
\begin{equation}\label{eq:rateCollest}
\begin{split}
\mathcal{R}_{DC}\approx
V(R_{col})n^2(R_{col})\pi(f_R R_\star)^2 v(R_{col})\\
\approx f^2_R\frac{1}{P_h}\left(\frac{\MBH}{m}\right)\left(\frac{R_\star}{R_h}\right)\;,
\end{split}
\end{equation}
where we used the BW relation $N(R_{col})=(R_{col}/R_h)^{5/4}N$ and also that by construction $N(R_h)\equiv\MBH/m$. For the Milky Way and sun-like values we repeatedly assume, this total estimate is $\mathcal{R}_{DC}\approx 10^{-7}\;\invyr$.

Notably, if DCs do indeed deplete the inner part of the cusp, stars are unlikely to evolve through GW emission in multiple inspiraling orbits all the way to $R_T$. The EMRI rate must therefore be reduced by the possibility of DCs. The TDE rate may also be affected, since some stars on eccentric orbits and $r\sim R_h$ could suffer collisions before finally being scattered into $J<J_{LC}$, but since only several orbits are required prior to the final plunge below $R_T$, this effect should be small. In any case, it is clear that collisions must be accounted for in any analysis of the rates of destructive processes in stellar cusps.

\section{The Monte Carlo Simulation}\label{sec:MCpluscollisions}

Despite continuous progress in $N$-body simulations (see, e.g., \cite{Nbody6pp}), tracking $\sim 10^6$ stars over long time scales is still impractical. This is especially true in our case, in which we need to follow stars close to the SMBH accurately where the periods are very short, while allowing the entire system to evolve over a cluster relaxation time. 

Instead, we opt for a simplified Monte Carlo algorithm. Each star is tracked individually in $(E,J)$ space, but the cumulative effect of other stars is averaged out according to the instantaneous density profile. The basics of such an algorithm were originally suggested by \cite{Henon1971}, and here we follow the technical approach presented by \cite{FragioneSari2018} and \cite{SariFragione2019}, to which we add a consistent treatment of stellar collisions. 

\subsection{Setup \label{subsec:setup}}

We consider a Milky Way-like cluster where $\MBH=4\times10^6\;\Msun$, which is surrounded by a cluster of $N=4 \times10^6$ identical stars with mass $m=1\;\Msun$. The cluster is assumed to extend up to a radius of influence of
\begin{equation}
R_h=\frac{GM}{\sigma^2}\approx 2\ \mathrm{pc}\ ,
\end{equation}
where $\sigma$ is the measured velocity dispersion external to the radius of influence \citep{MerrittFerrarese2001}. Stars occupy the range between the tidal radius $R_T=1.5\times 10^{13}\;\mathrm{cm}$ ($1\;$au) and $R_h$. 

At the beginning of the simulation, all stars are assigned values of the sma, $r$, and specific angular momentum, $J$. The sma is drawn randomly from a weighted distribution that produces a BW profile of $n(r)\sim r^{-7/4}$, and the specific angular momentum is set by uniformly sampling $(J/J_C(r))^2$  over the range 0 to $J_C(r)=\sqrt{GMr}$ (the circular angular momentum given the star's sma, $r$). The latter is essentially a thermal distribution, which generates an isotropic velocity field. Once $r$ and $J$ are set, the specific energy, $E$, period, $P$, eccentricity, $e$, and periapse, $r_p$, are determined for each star assuming Keplerian orbits
\begin{equation}\label{eq:properites}
\begin{split}
E=\frac{G\MBH}{2r},\;\;\;P=2\pi\left(\frac{r^3}{G\MBH}\right)^{1/2}\\
e^2=1-\frac{J^2}{G\MBH r},\;\;\;
r_p=(1-e)r\;.
\end{split}
\end{equation}

\subsection{Evolution by effective two-body scatterings \label{subsec:T2BJtime}}

Even though $r$ and $r_p$ are continuous and are stored individually for each star, it is numerically convenient to divide the $(r,r_p)$ plane into evenly spaced logarithmic bins. We set $r_{i+1}=2r_i$, $r_{p,j+1}=2r_{p,j}$, so $i$ and $j$ serve as indices of the resulting two-dimensional grid. Following \cite{SariFragione2019}, 
we track at each time step the number of stars in any particular bin, $N(i,j)$, and the total number of stars with an sma enclosed in a specific bin, regardless of periapse
\begin{equation}
\label{eq:N(i)}
N(i)=\sum_j N(i,j)\;.
\end{equation}

For each bin $(i,j)$, we calculate the momentary eccentricity-dependent time step,
\begin{equation}\label{eq:T2BJij}
\Delta\ttbj(i,j)=\frac{F}{N(i)}\frac{r_{p,j}}{R_h}\sqrt{\frac{r_i}{R_h}};,
\end{equation}
where 
\begin{equation}\label{eq:Fdef}
F=f_t\frac{P(R_h)(\MBH/m_\star)^2}{\gamma ln {\Lambda}}
\end{equation}
is a universal parameter of the system. We set $\ln\Lambda=10$, and the control parameter $\gamma=1.5$ that takes into account the bin size \citep{FragioneSari2018}, and also $f_T=0.1$ as the fraction of the relaxation time allowed in a single evolutionary time step.

Time in the simulation is advanced in discrete time steps, $\Delta t$, which are determined at each time $t$ by the fastest evolving grid point  $(i,j)$
\begin{equation}\label{eq:dt}
\Delta t=\min_{i,j} \Delta \ttbj(i,j)\ .
\end{equation}

Updating all stars at every time step is very expensive numerically. We use the scheme suggested by \cite{SariFragione2019} in which for each grid point $(i,j)$ we store the last time it was updated, $t_L(i,j)$, and update it again only when the current time, $t$, satisfies
\begin{equation}
\label{eq:update}
t-t_L(i,j)>\Delta \ttbj(i,j)\;.
\end{equation}

If a grid point $(i,j)$ is due to be updated, each individual star in that bin is allowed to evolve its orbital parameters. 
For each star, $k$, we identify the range of distances from the SMBH that it samples along its orbit, which correspond in the $i$-th component of the grid to the range $[W_1,W_2]$. For each bin $W$ we assume that the inspected star follows a random walk by a series of scatterings with the stars in that bin. The cumulative effect has an average impact parameter 
\begin{equation}\label{eq:Bimpact}
B(W)=r_W \sqrt{\frac{P_{*k}}{N(W)T_{2B}^{J}(i,j)}}\ ,
\end{equation}
where $P_{*k}$ is the period of the inspected star. The impact parameter is converted to typical changes in the specific energy and angular momentum due to the stars at grid cell $i=W$, with 
\begin{equation}\label{eq:DeltaE}
\Delta E(W)=\frac{Gm}{B(W)}\;,
\end{equation}
and
\begin{equation}\label{eq:DeltaJ}
\Delta J(W)=r_W\frac{Gm}{v(W)B(W)}\;,
\end{equation}
where $v(W)=\sqrt{G\MBH/r_W}$ is the characteristic velocity at $r_W$.
 Finally, the energy and the angular momentum of the star are updated with respect to their current values of $E_{old}$ and $J_{old}$ to
\begin{eqnarray}
\label{eq:step2BE}
E_{new}&=&E_{old}+\sin{\chi} \Delta E \\
\label{eq:step2BJ}
J_{new}&=&\left({J_{old}^2+\Delta J^2-2 J_{old}\Delta J \cos \Phi}\right)^{1/2}\ ,
\end{eqnarray}
with $0\le \chi < 2\pi$ and $0\le \Phi < 2\pi$ randomly drawn from a uniform distribution. The star's orbital properties, $r$,$P$, $r_p$, and $e$, are then updated consistently. This process is repeated for the entire range $[W_1,W_2]$ for the inspected star, and for all stars in the $(i,j)$ bin that is being advanced in time.
 
It is worthwhile to point out explicitly the codependency between the choice of grid separation ($r_{i+1}/r_i$) and the approximations used in Equations (\ref{eq:Bimpact}-\ref{eq:DeltaJ}). The latter assumes that two-body interactions are dominated by multiple small scatterings at large distances. These assumptions break down for thin spatial bins ($r_{i+1}/r_i\ll2$). Not only will the assumption $N(r)\gg 1$ be affected, but also the identification implied in equation \ref{eq:Bimpact}, that a star contributes to scattering other stars only in the bin that includes its sma. This identification is appropriate, of course, for roughly circular orbits, but with $r_{i+1}/r_i=2$ it is actually consistent even with eccentric orbits and an anisotropic velocity distribution: practically all stars do indeed spend most of their period in a single spatial bin, and can be approximated as such in terms of their contribution to the average gravitational scattering. This last point is especially relevant to stars which are captured following a binary disruption (see below). In summary, given the nature of the approximations, increasing the spatial resolution does not necessarily improve the accuracy of the results. These shortcomings are best addressed by direct $N$-body simulations of the problem, which we report separately (G. Yassur and S. Balberg 2023, in preparation).     

\subsection{Evolution by GW emission \label{subsec:TGR}}

In each time step, stars evolve their orbits due to GW emission. The changes in energy and angular momentum are calculated with standard formulae \citep{Peters1964,HopmanAlexander2005}
\begin{eqnarray}
\label{eq:stepGWE}
E_{new}&=&E_{old}+\Delta E_{GW} \\
\label{eq:stepGWJ}
J_{new}&=&J_{old}+\Delta J_{GW}\ ,
\end{eqnarray}
where
\begin{eqnarray}
\Delta E_{GW}&=&\frac{8\pi}{5\sqrt{2}}f(e)\frac{m c^2}{\MBH}\left(\frac{r_p}{r_S}\right)^{-7/2} \frac{\Delta t}{P(r)} \\
\Delta J_{GW}&=&-\frac{16\pi}{5}g(e)\frac{Gm}{c}\left(\frac{r_p}{r_S}\right)^{-2} \frac{\Delta t}{P(r)}\ ,
\end{eqnarray}
and
\begin{eqnarray}
f(e)&=&\frac{1+(73/24)e^2+(37/96)e^4}{(1+e)^{7/2}} \\
g(e)&=&\frac{1+(7/8)e^2}{(1+e)^2}\ .
\end{eqnarray}
For simplicity, we follow \cite{SariFragione2019} and limit GW evolution only to stars with $r_p\leq r_{p,GR}$, where we set $r_{p,GR}=0.1R_h$. 

\subsection{TDEs, EMRIs, and Ejections\label{subsec:TDE-EMRI-Eject}}

When the new angular momentum of a star corresponds to an orbit with $r_p\leq R_T$, it is considered to be tidally destroyed by the SMBH. If this occurs following an effective two-body step (equation \ref{eq:step2BJ}) the event is registered as a TDE, and if following a GW step (equation \ref{eq:stepGWJ}) the event is registered as an EMRI. 

Since our goal is to study the steady-state cusp profile and rates of destructive events, we set to maintain a constant number of stars. We assume that every destroyed star is replaced by a new one that has been scattered into the cusp from the $r>R_h$ cluster, which obviously is not accounted for in our simulation. When stars from disrupted binaries are not included (see below), replacement stars are randomly generated in the outer bin of $r$ (weighted to maintain an $n(r)\sim r^{-7/4}$ profile in this bin), and again with a thermal distribution of eccentricities. 

Another possible outcome of an effective two-body step is that a star scatters to energy $E<G\MBH/2R_h$, indicating that that star was ejected from the $r\leq R_h$ cusp. Again, attempting to maintain a steady-state cusp, we return the star to its previous $(E_{old},J_{old})$ quantities. We essentially assume that every ejected star is replaced by a star with identical properties that has been scattered into the cusp from the $r>R_h$ cluster. 

\subsection{Destructive collisions \label{subsec:calccollisions}}

Our principal innovation in this work is the inclusion of destructive two-star collisions in the inner part of the cusp. For the purposes of the simulation, we set that DCs are possible at distances smaller than $R_{col}=2.4\times 10^{16}\;$cm ($\sim 8\;$mpc), where the typical kinetic energy of a single star is about $10^{49}\;$erg, exceeding the binding energy of a Sun-like star. We thus neglect destruction through multiple low-velocity collisions at larger distances \citep{Roseetal2023}. These will certainly affect the stellar population, but are not relevant as sources of an energetic optical flare in a single event.

Our procedure for tracking collisions is as follows. We divide the radial positions from the SMBH into bins, and for convenience we use the same grid spacing as for the smas. Physical space is then divided into spherical shells with volume $V_i=4\pi(r^3_{i+1}-r^3_i)/3$.

For each star $k$ that has an orbit with $r_p<R_{col}$, we calculate the probability $p_{*k}(i)$ to find the star in the volumetric bin $i$. The simplest cases are, $p_{*k}(i)=1$, if the orbit is entirely contained in a single spherical shell, or $p_{*k}=0$, if the star does not enter the volumetric bin $i$. For an orbit that crosses in and out 
of a bin, the probability is calculated through the fraction of time that the star spends in the shell out of the entire orbit. Denoting the time a star spends inside the spherical shell $i$ as $\Delta t_{*k}(i)$, the probability of finding the star in the $i$ spherical shell is
\begin{equation}\label{eq:pstark}
p_{*k}(i)=\frac{\Delta t_{*k}(i)}{P_{*k}}=\frac{2}{P_{*k}}\int^{x_{max*k}(i)}_{x_{min*k}(i)}\frac{dx}{v_{r*k}(x)}\;,\;.
\end{equation}
The integral in Equation (\ref{eq:pstark}) includes $v_{r*}(x)$, the radial component of the star's velocity when at $x$, which for Keplerian motion with with sma $r_{*k}$ and eccentricity $e$ is 
\begin{equation}\label{v_r(x)}
v_{r*k}(x)=\left(\frac{2G\MBH}{x}-\frac{G\MBH}{r}-\frac{G\MBH r}{x^2}(1-e^2)\right)^{1/2}\;.
\end{equation}

The limits on the integral in Equation (\ref{eq:pstark}) can be $x_{min*k}(i)=r_i$ and $x_{max*k}(i)=r_{i+1}$ if the orbit passes through both surfaces of the shell $(i)$, or $x_{min*k}(i)=r_{p*k}$ or $x_{max*k}(i)=r_{a*k}$ if the star only partially penetrates the shell, where $r_{p*k}$ ($r_{a*k}$) is the star's periapse (apoapse). 

The vector $p_{*k}$ is converted into an effective, time-averaged, density of stars in each spherical shell
\begin{equation}\label{eq:neff(i)}
n_{eff}(i)=\frac{\sum_k p_{*k}(i)}{V(i)}\;.
\end{equation}
This quantity is used to calculate an effective optical depth for collisions in the $i$th shell. We denote the total distance the $k$-th star travels though the shell in a single orbit as $\Delta x_{*k}(i)$. It may vary from $2\pi r_{*k}$ for circular orbits to $2(x_{max*k}(i)-x_{min*k}(i))$ for very eccentric orbits. The optical depth the $k$th star experiences in the $i$th spherical shell is then  
\begin{equation}
\label{eq:tauki}
\tau_{*k}(i)=\left(n_{eff}(i)-\frac{p_{*k}(i)}{V(i)}\right)\times\pi (f_R R_\star)^2\Delta x_{*k}(i)\;.
\end{equation}
We remark that in Equation (\ref{eq:tauki}) the weight of the specific star $k$ is subtracted from the effective density, in order to remove the fictitious option of the star colliding with itself. 

Finally, we calculate the probability that the $k$th star will collide with another star during a single orbit is 
\begin{equation}\label{eq:pnocolki}
p_{col,*k}=\frac{1}{2}\left[1-e^{-\tau_{*k}}\right];\;\;  
\tau_{*k}=\sum_i \tau_{*k}(i)\;.
\end{equation}
Note that the factor $1/2$ in the left-hand part of Equation \ref{eq:pnocolki} is applied as an effective correction, appropriate for small $\tau$, in order to avoid double counting each possible collision when the vector $p_{col,*k}$ is compiled over all stars. The probabilities $p_{col,*k}$  are typically minuscule, given the very small size of the stars.

The values of $p_{*k}(i)$ and $p_{col,*k}$ are calculated at the beginning of each evolutionary time step. 
However, applying them in the calculation presents technical challenges, due to the very wide range of time scales involved. First, the periods of  the candidate stars for collisions vary over more than four orders of magnitude. Second, in our evolutionary scheme the time step size, $\Delta t$, is set by the stars with the shortest $\ttbj$ (equation \ref{eq:T2BJij}). These are typically the stars with sma $r\approx R_h$ and the highest eccentricities. The  evolutionary time step is therefore of order $(R^3_h/(G\MBH))^{1/2}$, which is several orders of magnitude greater than the orbital periods of the inner stars. Tracking collisions per orbit per star is computationally prohibitive. 

We overcome this complexity by estimating the probability for a collision per star over multiple orbits. The probability that a star will {\it not} suffer a collision over the entire evolutionary time step $\Delta t$ is
\begin{equation}\label{eq:Pnocol}
p_{nocol,*k}(\Delta t)=(1-p_{col*k})^{\Delta t/P_{*k}}\;.
\end{equation}
Even though $\Delta t/P_{*k}$ is large, $p_{col,*k}$ is typically very small. The result, $p_{nocol,*k}(\Delta t)$, is almost always very close to unity, and is readily evaluated. 

Equation \ref{eq:Pnocol} includes a nontrivial assumption, which is that each star randomizes the orientation of its orbit over a single period. Obviously this assumption is incorrect in Newtonian dynamics, but is more realistic when considering relativistic precession, which is relevant in the inner cusp close to the SMBH, where we attempt to estimate the collision rate. A single simulation time step is much longer than an orbital period (and both are much shorter timescales than $T_{GW}$ which evolves the orbit in terms of $(r,r_p)$), so randomization appears to be a crude, but reasonable, assumption. We examine this point separately in specific N-body simulations (G.~Yassur and S.~Balberg, in preparation).  
Given the final vector $p_{nocol,*k}(\Delta t)$, we scan at the beginning of each evolutionary time step all the stars with $r_p\leq R_{col}$. For each star we draw from a random number $0<p_0<1$, and in the rare cases in which $p_0<1-p_{nocol,*k}(\Delta t)$ the star is decreed as having experienced a collision. If this happens, the simulation determines the spherical shell where the collision happened, by drawing from the weighted distribution of $exp[-\tau_{*k}(i)]$ (Equation \ref{eq:tauki}). 
The partner star is determined by drawing from the weighted distribution of $p_{*k}(i)$ (Equation \ref{eq:pstark}) in the determined $i$th shell. The collision is registered, along with its location in terms of the index $i$ of the spherical shell, and the two collided stars are removed from the simulation. Again, since we seek a steady state, the removed stars are replaced in a similar fashion to those lost in TDEs and EMRIs, as described in section \ref{subsec:TDE-EMRI-Eject}. 

We comment that in principle a new star can itself have a nonzero probability for collision. This is a very rare possibility, but in order to account for it we examine a secondary step, in which we calculate $p_{col*k}$ for the replacement stars, and check for the probability of them colliding with other relevant stars. This is done for a shorter time $\Delta t^\prime<\Delta t$, drawn exponentially from $0<\Delta t^\prime/\Delta t<1$ (since we do not compute the actual time at which the first collision took place). We find that this precaution is basically meaningless when stars from disrupted binaries are ignored, while when they are included (Section \ref{sec:Resultswithbinaries}) it does generate a small correction to the results. 

It is noteworthy that an initial setup according to the BW profile does create relatively large collision probabilities. For such a profile several collisions occur during a time of order $P(R_h)$. Obviously, this is an artifact of the steep density gradient of the BW profile, which does not account for collisions. To remedy this conflict, we processes the initial stellar distribution setup (Section \ref{subsec:setup}) with the collision algorithm described above once for time step $\Delta t_0$ prior to actually starting a time-dependent evolution. We set  $\Delta t_0=(R^3_h/G\MBH)^{1/2}$ and remove and replace stars that are determined to have collided (typically of order 10, depending on the choice of $f_R$). These collisions are not included in the total count of collisions which occur during the simulation.

\section{Results Part I: Destructive rates without stars from disrupted binaries}\label{sec:Resultsnobinaries}

We begin by examining the impact of stellar collisions on the rates of destructive events and the cusp structure when stars from disrupted binaries are not considered. The significance of adding such stars is discussed below in Section \ref{sec:Resultswithbinaries}.

\begin{figure}
    \centering
    \includegraphics[width=\columnwidth]{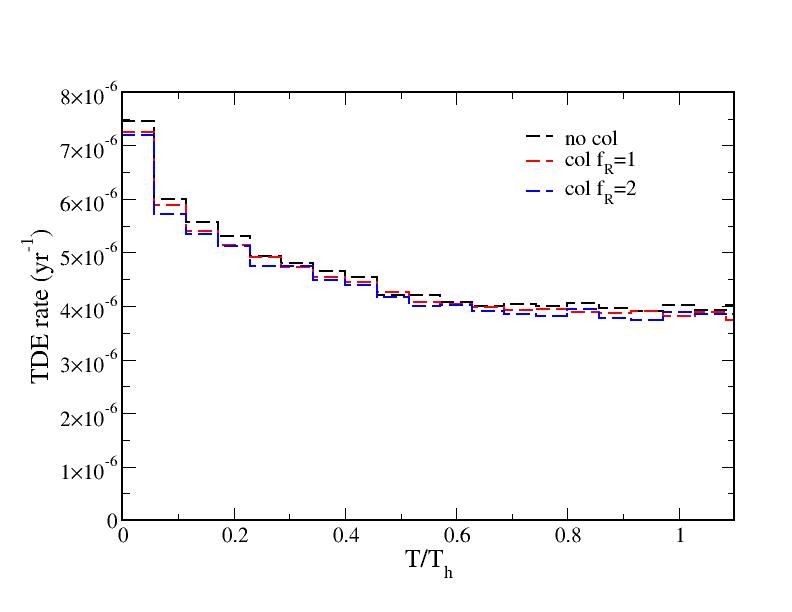}
    \includegraphics[width=\columnwidth]{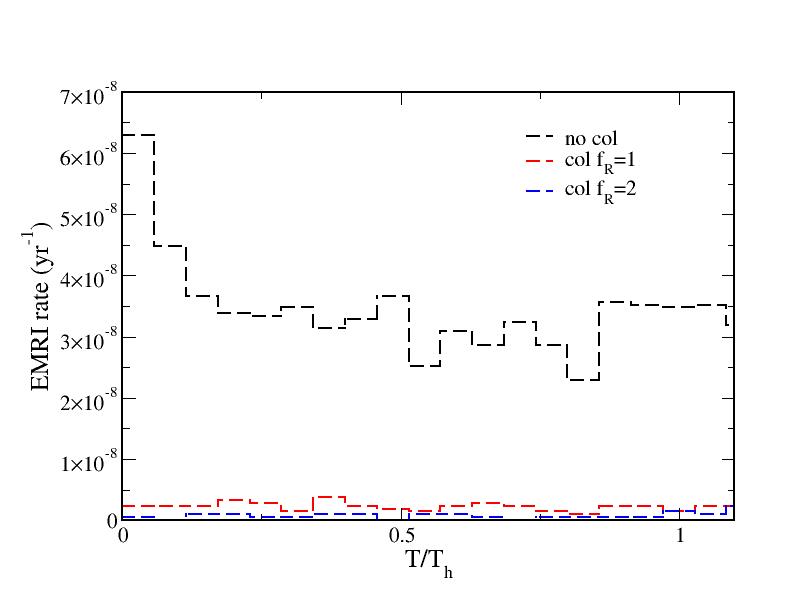}
    \includegraphics[width=\columnwidth]{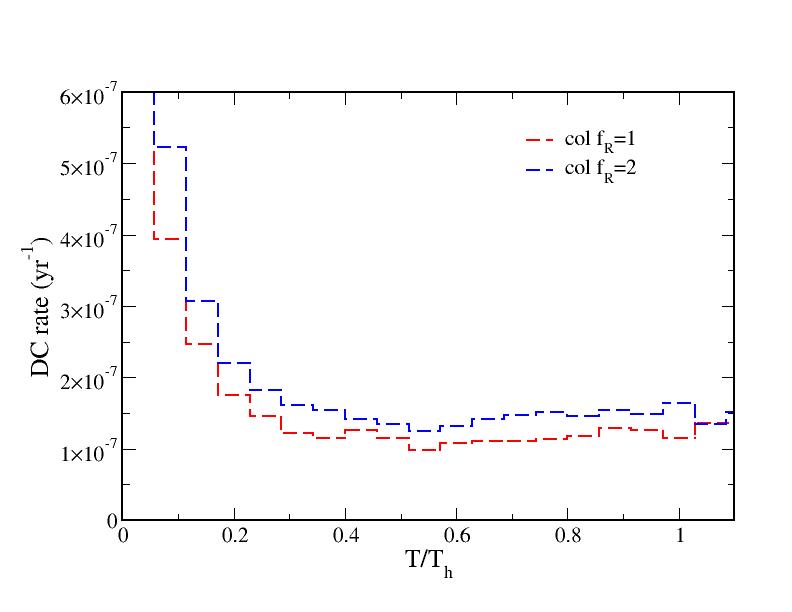}
    \caption{Top to bottom: the TDE, EMRI, and DC rates, respectively, as a function of time with and without allowing for destructive stellar collisions. The collision cross section is taken as 
    $\pi (f_R R_\star)^2$; setting $f_R=0$ corresponds to no collisions. Notably, including collisions hardly affects the TDE rate, but suppresses EMRIs almost completely.} 
    \label{fig:Allrates_no bin}
\end{figure}

In figure \ref{fig:Allrates_no bin} we show the calculated TDE, EMRI, and DC rates as a function of time. We ran the simulations for approximately one cluster relaxation time, $T_h$ (equation \ref{eq:T2B} for $r=R_h$). Notably, the system roughly settles on a steady state after a few tenths of $T_h$ (since the simulation is initialized with a BW $n(r)\propto r^{-7/4}$ cusp that does not account for GW emission or stellar destruction).  

When collisions are not accounted for, we generally reproduce the qualitative and quantitative results found by \cite{SariFragione2019}. Specifically, we find that the steady-state TDE rate is about $4\times 10^{-6}\invyr$, in good agreement with the order-of-magnitude estimate presented above in Section \ref{subsec:twobodyNR}. The steady-state EMRI rate settles on about $3.5\times 10^{-8}\invyr$, or about $1\%$ of the TDE rate, again in good agreement with the estimate in Section \ref{subsec:GW}. 

Figure \ref{fig:Allrates_no bin} also includes the rates when DCs are taken into account. Shown are the results for both the conservative estimate $f_R=1$ and the upper limit $f_R=2$. We see that the TDE rate is almost unaffected by the collisions. This is to be expected, since DCs occur close to the SMBH, while the dominant source of TDEs are stars in the vicinity of $R_h$ which plunge into the SMBH after being scattered into the loss cone. These are scattered into an orbit with $J\leq J_{LC}$ over a time scale of a few $P(R_h)$, so most TDE candidate stars sample the collision-prone region of the cusp only a few times before being disrupted. 
This insight is consistent with the DC rates presented in the bottom panel in Figure \ref{fig:Allrates_no bin}. These rates tend to settle on a steady-state value of slightly over $10^{-7}\invyr$, in very good agreement with the estimate presented in Equation \ref{eq:rateCollest}; DCs are mostly due to stars which diffuse in the $r\leq R_{col}$ region gradually and spend most of their orbits in the inner cusp. The fraction of TDE candidates that were lost to collisions is small.   

The effect on the EMRI rate, on the other hand, is dramatic. Collisions essentially deplete the inner part of the cusp, and since stars can collide prior to any orbital evolution, EMRIs are suppressed almost completely. The EMRI rate drops to few times $10^{-9}\invyr$, $10\%$ or less of the estimated rate without collisions.

Note that the steady-state value of $\mathcal{R}_{DC}$ is essentially insensitive to the value of $f_R$ (as long as it is nonzero). Collisions dominate destruction in the inner cusp, and so simply balance the rate at which stars are supplied to the region $R\leq R_{col}$, which is a general property of the cluster. Correspondingly, the density profile in the inner cusp adjusts so that integrating over $(n(r)f_R)^2$ at every distance $r$ from the SMBH yields the appropriate collision rate. Setting $f_R=2$ mostly reduces the steady-state densities by about a factor of 2 compared to the $f_R=1$ case. The enhanced depletion of $n(r)$ for $f_R=2$ does have a nonnegligible effect on the EMRI rate, since it is proportional to the number of stars in the inner region of the cusp. But this is of secondary importance, given that $\mathcal{R}_{EMRI}$ is very small. 

\begin{figure}
    \centering
\includegraphics[width=\columnwidth]{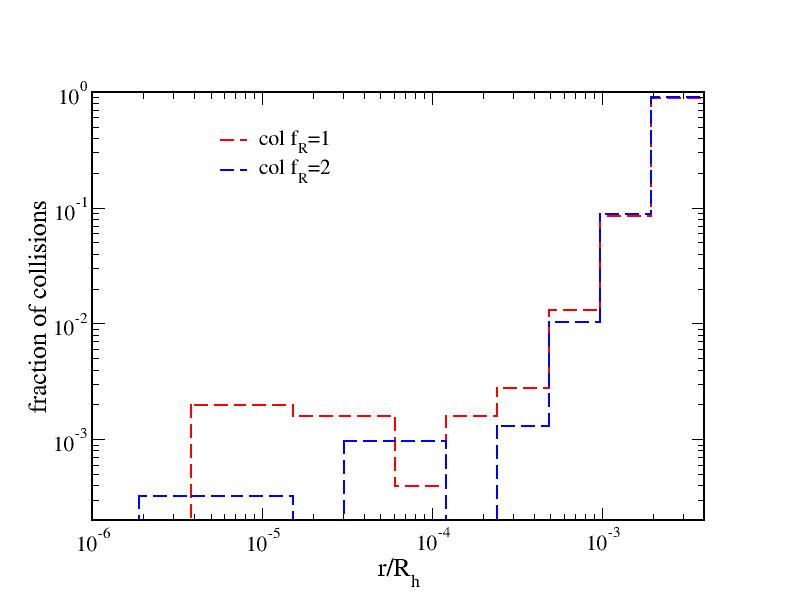}
    \caption{The fraction of collisions as a function of distance from the SMBH out of the total number of collisions, averaged over the later half of the simulations.}
    \label{fig:Colgrid_nobin}
\end{figure}

Figure \ref{fig:Colgrid_nobin} shows the fraction of collisions that occurred in the different volumetric bins. In order to asses the steady state distribution of collisions, these fractions are averaged only over the later half (time wise) of the simulation. 
Quantitatively, we find that about $99\%$ of collisions occur at distances $\gtrsim 3\;$mpc from the SMBH. This is, of course, consistent with the total DC rate found above, recalling that most stars that are supplied into the $r\leq R_{col}$ region of the cusp evolve first into an $r\lesssim R_{col}$ orbit. The total energy per collision at $R_{col}$ is by construction of order $10^{49}\;$ erg. The implications on the potential observability of such collisions are dire: the thermal energy in the collisions will be small and the optical flare will be very dim. Some of the ejected gas may eventually accrete onto the SMBH, but given the large initial distance, the accretion rate will be low, leading to a prolonged, low-luminosity event, if at all. This last result changes dramatically when stars from disrupted binaries are included in the simulations, as we demonstrate below.  

\section{Results Part II: destructive rates including stars from disrupted binaries\label{sec:Resultswithbinaries}}

As mentioned above, disrupted binaries add a unique channel for transporting stars to inner regions of the stellar cusp. Generally, a binary with separation distance $a_b$ is disrupted when it approaches the SMBH closer than its tidal disruption radius, $R_{Tb}(a_b)$ \citep{Sarietal2010}
\begin{equation} \label{eq:R_Tb}
R_{Tb}(a_b)\approx \left(\frac{\MBH}{m}\right)^{1/3}a_b\;.
\end{equation}

As is the case for single stars, binaries are far more likely to arrive at their tidal disruption radius after being scattered into an elongated orbit around the SMBH rather than by diffusing inward through a series of quasi-circular orbits. As a result, a disrupted binary typically has an original total energy close to zero, and once disrupted the three-body process with the SMBH often results in one star being ejected with positive energy \citep{Bromleyetal2006,PeretsSubr2012}, with the other being captured in a tight orbit \citep{Rossietal2014}. 
Indeed, ejection through binary disruption is considered the favored production mechanism for the hyper-velocity-Stars (HVS) which are observed leaving the plane of the Milky Way with velocities exceeding $1000\;{\rm km\; s^{-1}}$ \citep{Brown2015}. The bound companion is left in a highly eccentric orbit around the SMBH. Its periapse is approximately equal to the binary tidal disruption radius, $r_p\approx R_{Tb}(a_b)$, and energy conservation requires an sma of approximately 
\begin{equation} \label{eq:R_TB}
r(a_b)\approx \left(\frac{\MBH}{m}\right)^{2/3}a_b=\left(\frac{\MBH}{m}\right)^{1/3}R_{Tb}(a_b)\;.
\end{equation}
For the Milky Way SMBH and Sun-like stars, this implies eccentricities of $e\approx0.99$.

Hereafter we refer to stars that were disrupted from binaries and obtain bound, eccentric orbits as "injected stars". Their contributions to the TDE, EMRI, and DC rates, as well as their impact on the stellar density profile, obviously depend on two parameters: the total rate, $\eta_b$, at which they are injected, and the distribution of injected stars over the $(r,r_p)$ space. The former depends on the fraction of stars which exist in binaries in the vicinity of $R_h$, while the latter reflects the original distribution function of binary separation distances.  

Some nontrivial assumptions are required to set the distribution of $a_b$, since it is a combined result of the inherent characteristics of binaries and how they can survive being disrupted by other stars \cite[i.e. "ionized", ][]{HeggieHut1993}. Tighter (smaller $a_b$) binaries should clearly be more resilient to disruption by multibody forces. According to \cite{Hopman2009}, the probability of a binary with a given value of $a_b$ to survive disruption rises rapidly the further it lies from the SMBH. For Milky Way parameters and Sun-like stars, binaries with $a_b\leq 0.2\;$au should survive disruption at $R_h$. On the other hand, tighter binaries have smaller tidal disruption radii, and therefore smaller loss cones in angular momentum space and a smaller probability of disruption per binary. Conversely, wider (larger $a_b$) binaries which can survive somewhat outside of $R_h$ also have larger tidal disruption radii, and so they too can be scattered into disruptable eccentric orbits close to the SMBH with comparable probabilities. 

A full analysis of the proper distribution of binary separation distances is beyond the scope of this work, and hereafter we limit ourselves to two simplified options. As a nominal model, we assume that the maximal value is $a_{max}=1\;$au (significantly wider binaries are both unlikely to survive in or close to the cusp, and will not contribute to the density in the $r\leq R_{col}$ region). We also consider the case $a_{max}=0.1\;$au, which essentially assumes that only very tight binaries exist at any region that serves as a source for injected stars. In any case, we set the lower limit of $a_{min}=0.01\;$au which is the minimal separation distance for two sun-like stars.  

\subsection{Including stars from disrupted binaries in the simulations \label{subsec:AddBin}}

Our algorithm for including stars from disrupted binaries is as follows. When a star is destroyed, a random number $0\leq\xi<1$ is uniformly drawn and compared to a predetermined parameter $p_b$. If $\xi>p_b$ the star is replaced by a new star in the outer sma bin (as discussed in \ref{subsec:TDE-EMRI-Eject}); setting $p_b=0$ means no injection of binaries. If $\xi\leq p_b$ a star is injected with the appropriate parameters from a disrupted binary: $r_p=r_{Tb}(a)\;$ and $ r=100r_p$. Note that this method conserves the total number of stars, but the steady-state rate of injected stars is in turn consequential, rather than a predetermined parameter (our method is thus differs from \cite{SariFragione2019}, who injected stars at a constant rate). For a given set of external parameters, we can fine tune the parameter $p_b$ in order to generate any specific steady-state rate $\eta_b$. 

The original binary sma $a_b$ is drawn according to a log-normal distribution 
\begin{equation}\label{eq:abspread}
f(a_b)\propto \frac{1}{a_b}
\end{equation}
which conforms with the observed population of binaries \citep{DucheneKraus2013}. This distribution is assumed to lie between two values, $a_{min}$ and $a_{max}$. As mentioned above, we set  $a_{min}=0.01\;$au, and consider two values for $a_{max}$: $1\;$au and $0.1\;$au. 

\subsection{The effects of binaries without collisions \label{subsec:BinNoCol}}
\begin{figure}
    \centering
\includegraphics[width=\columnwidth]{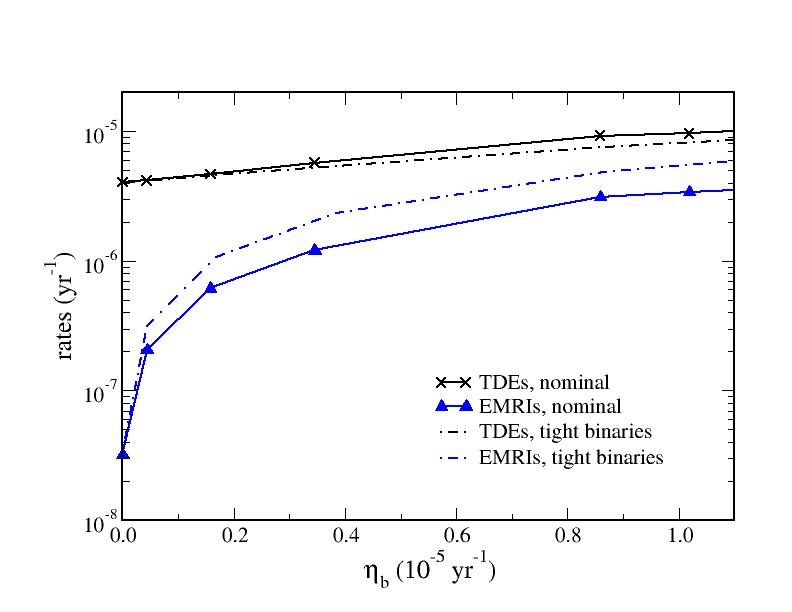}
    \caption{Steady-state TDE and EMRI rates as a function of the injection rate of stars from disrupted binaries. Collisions are ignored. The rate $\eta_b$ is generated by varying the probability $p_b$, where stars destroyed in a TDE or an EMRI are replaced with an injected star with probability $p_b$, and by a star at $R_h$ with probability $1-p_b$.  Shown are the results for the nominal assumption, $a_{max}=1\;$au and for the tight binary assumption, $a_{max}=0.1\;$au.}
    \label{fig:TDE+EMRIs_bin_NOcol}
\end{figure}

Figure \ref{fig:TDE+EMRIs_bin_NOcol} presents the results of the simulations when collisions are ignored, in terms of the steady-state values of $\mathcal{R}_{TDE}$ and $\mathcal{R}_{EMRI}$,
as a function of the steady-state injection rate, $\eta_b$. Again, we clarify that $\eta_b$ is itself a result of the simulation, which we fine tune (by trial and error) by setting the parameter $p_b$. 
 
We find that inclusion of injected stars from disrupted binaries increases both the TDE rate and the EMRI rate. The results are consistent with those of  
\cite{SariFragione2019}. Given the small initial periapses and high eccentricities, injected stars have a much larger probability (compared to stars that enter the cusp at $R_h$) to be scattered into orbits that terminate in destruction at $R_T$ (see \citet{Bromleyetal2012}). Numerically, we find that about $70\%$ of the stars which were injected over the second half of the simulation were destroyed, in very good agreement with the analytical estimate of \cite{FragioneSari2018}. The majority of injected stars are destroyed as TDEs, but a significant minority inspiral to $R_T$ through the EMRI channel, either because they are initially injected into a combination of $(r,r_p)$ for which GW emission leads to an EMRI, or because they are quickly scattered into such a combination. Hence, both $\mathcal{R}_{TDE}$ and $\mathcal{R}_{EMRI}$ are enhanced by an addition of order $\eta_b$, which for the latter can result in an increase of over an order of magnitude in comparison to the case of no binary disruptions.
In particular, setting $p_b=0.81$ yields a corresponding injection rate of $\eta_b\approx 10^{-5}\invyr$, and results in steady state-rates of $\mathcal{R}_{TDE}\approx 9\times 10^{-6}\invyr$ and $\mathcal{R}_{EMRI}\approx 3\times 10^{-6}\invyr$. 

\begin{figure}
    \centering
\includegraphics[width=\columnwidth]{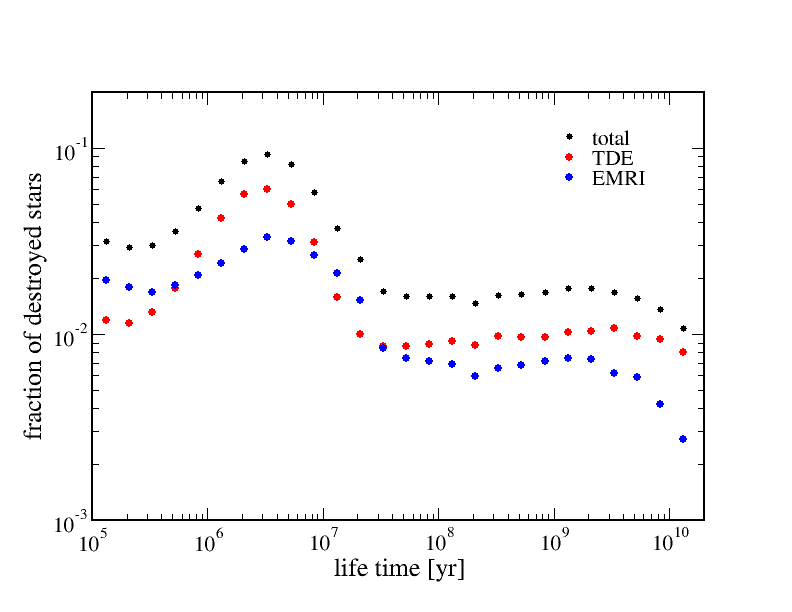}
    \caption{The distribution of lifetimes experienced by the injected stars, for simulations with $\eta_b=10^{-5}\;\invyr$, and for the nominal value $a_{max}=1\;$au. Shown are the total fraction and separately the fractions for stars destroyed as a TDE and stars destroyed as an EMRI. Collisions are ignored.}
    \label{fig:lifetime_binNOCOL}
\end{figure}

An interesting quantity of the destroyed injected stars is the lifetime they experience in the system. Figure \ref{fig:lifetime_binNOCOL} presents the distribution function of the lifetimes of injected stars that were destroyed during the later half of the simulation. We emphasize that these are only injected stars, not stars from the original stellar profile or stars added to the cusp at $R_h$. The total function, as well as the breakdown between the TDE channel and the EMRI channel are displayed. We comment that a few percent of the stars are destroyed within one time step of their injection (typically if injected very close to the SMBH), for which the simulation does not have the appropriate temporal resolution. These stars are not depicted in the figure.

We see that the lifetimes are spread over the entire range of a few times $10^5\;$ to $10^{10}\;$yr (essentially the half-length of the simulation). The long-lived stars are generally scattered into more circular orbits and then merge with the preexisting population, for which there is a finite probability of being rescattered toward destruction.
On the other hand, there is a maximum likelihood for lifetimes of a few $10^6\;$yr, both in TDEs and in EMRIs. This reflects the fact that the surplus of stars over the range of injection radii (compared to a BW profile) shortens the two-body scattering time scale (equation \ref{eq:T2BJ}). As a result, the injected stars, which inherently have large eccentricities, evolve faster than expected in a BW profile, so that a large fraction reach $r_p=R_T$ either by two-body scatterings or by GW emission.  

\subsection{Including binaries and collisions \label{subsec:BinCol}}

\begin{figure}
    \centering
    \includegraphics[width=\columnwidth]{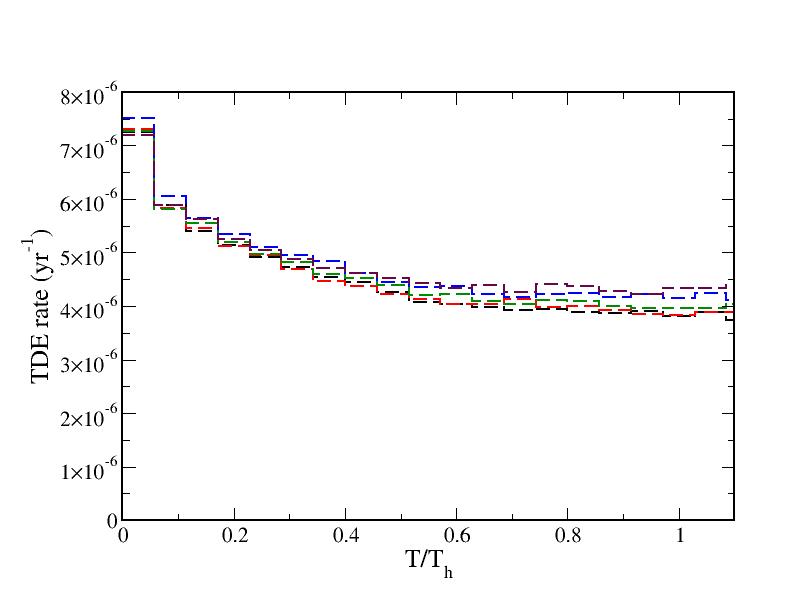}
    \includegraphics[width=\columnwidth]{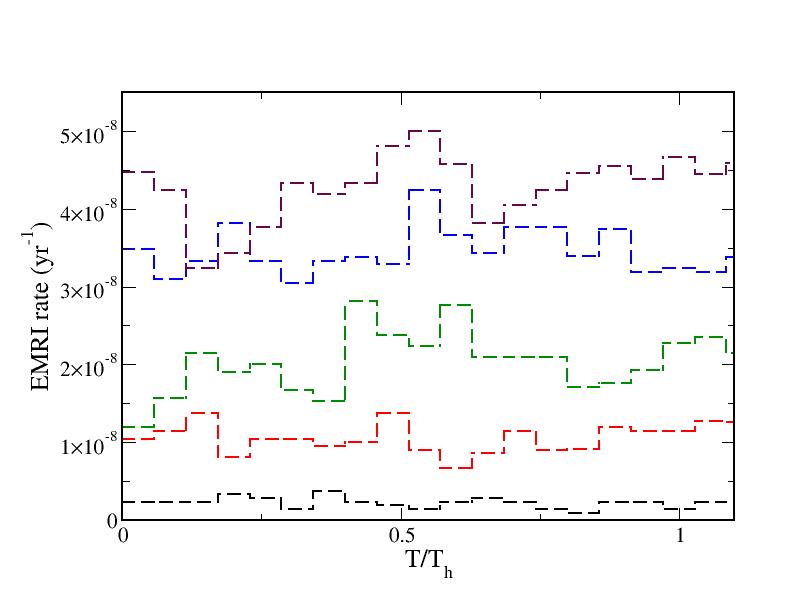}
    \includegraphics[width=\columnwidth]{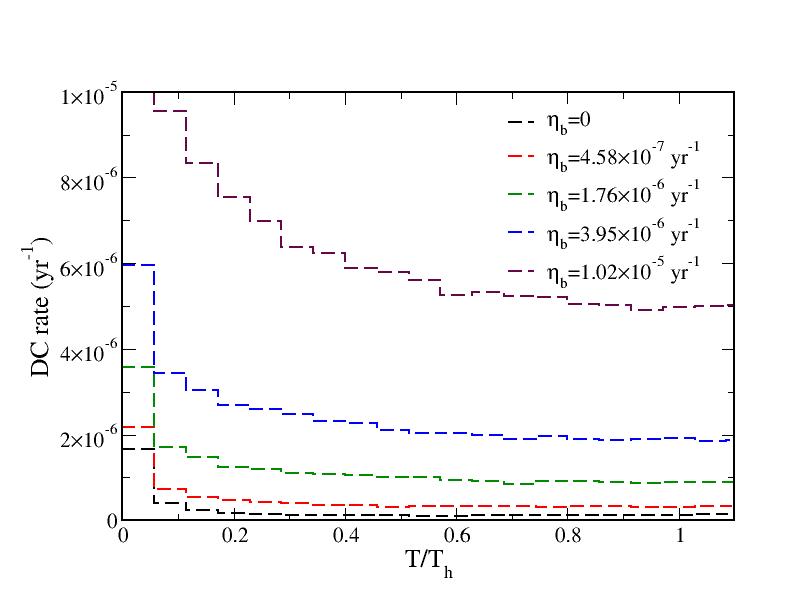}
    \caption{Top to bottom: the TDE, EMRI, and DC rates, respectively, as a function of time, for various values of the injection rate, $\eta_b$. Different curves correspond to different injection rates, noted in the legend. The collision cross section is taken as 
    $\pi (f_R R_\star)^2$, with $f_R=1$, and the injected stars are drawn from a log-normal distribution with $a_{max}=1\;$au (see Section \ref{subsec:AddBin}).} 
    \label{fig:Allrates_bin}
\end{figure}

Our results change quantitatively when collisions are taken into account. In Figure \ref{fig:Allrates_bin} we show the calculated TDE, EMRI, and DC rates as a function of time, again over approximately one cluster relaxation time, $T_h$. All simulations have the nominal values of $f_R=1$ for collisions and $a_{max}=1\;$au for the distribution of  the $(r,r_p)$ combinations for the injected stars. In the different simulations we varied the constant $p_b$, generating a different steady-state injection rate, $\eta_b$. Similar to Section \ref{sec:Resultsnobinaries}, the system roughly settles on a steady state after a time of a few tenths of $T_h$.  

\begin{figure}
    \centering
\includegraphics[width=\columnwidth]{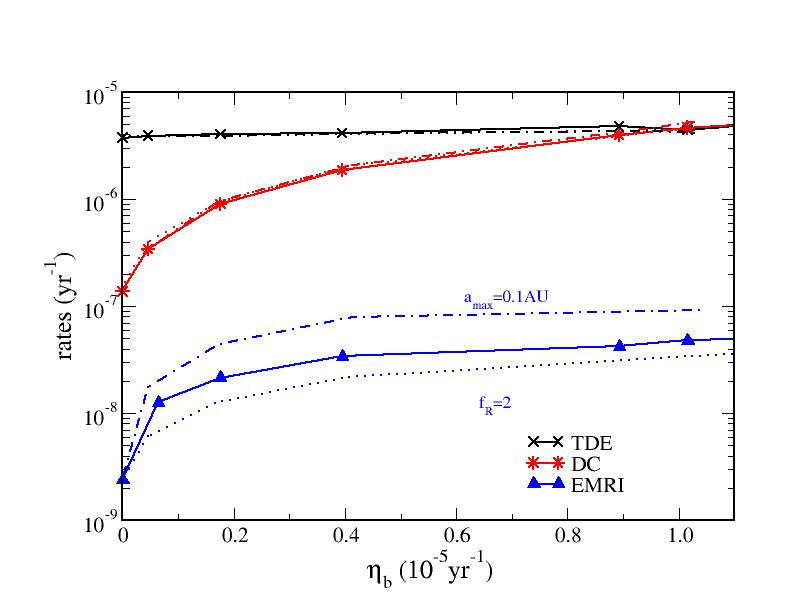}
    \caption{TDE (black), EMRI (blue), and DC (red) rates as a function of the injection rate of stars from disrupted binaries, $\eta_b$. Shown are the results for the nominal combination $f_R=1$ and $a_{max}=1\;au$, and also for changing (separately) to $f_R=2$ and $a_{max}=0.1\;$au. Note that these changes have a noticeable effect only on $\mathcal{R}_{EMRI}$.}
    \label{fig:TDE+DC+EMRIs_bin_colnom}
\end{figure}

As is to be expected, increasing the injection rate means more stars are supplied to the $r\leq R_{col}$ region, and so significantly more DCs and EMRIs, but only a minor effect on the TDE rate is seen. In this case as well DCs dominate over EMRIs by more than an order of magnitude. This result is directly depicted in Figure \ref{fig:TDE+DC+EMRIs_bin_colnom}, which  presents the results of the simulations in terms of the steady-state values of $\mathcal{R}_{DC}$, $ \mathcal{R}_{TDE}$ and $\mathcal{R}_{EMRI}$  as a function of the steady-state injection rate, $\eta_b$. 

We find that the inclusion of stars from disrupted binaries causes a dramatic increase of the collision rate, when compared to the same rate when binaries are not included. For low injection rates the new stars combine with those supplied by the cusp, creating larger densities and a significant growth of $\mathcal{R}_{DC}$. For higher values of $\eta_b$ collisions become the dominant destruction mechanism in the inner cusp; they self-regulate the density profile, and the slope of the $\mathcal{R}_{DC}(\eta_b)$ curve is moderated to a roughly linear one. Nonetheless, for the higher range of $\eta_b\gtrsim 5\times 10^{-6}\invyr$, $\mathcal{R}_{DC}$ is enhanced by more than an order of magnitude (compared to the value at $\eta_b=0$), and becomes comparable to $\mathcal{R}_{TDE}$. The latter is now essentially a constant, limited almost entirely to stars plunging from $R_h$ on eccentric orbits. 

Note that as in Section \ref{subsec:BinNoCol}, $\mathcal{R}_{EMRI}$ is also enhanced by a factor of $\sim 10$ for the higher end of $\eta_b$ values. However, since the evolutionary paths to EMRIs are still significantly suppressed by collisions, their total rate remains small. The enhancement of $\mathcal{R}_{EMRI}$ due to injected stars is of order $10^{-2}\eta_b$, rather than by $\sim\eta_b$, the result when collisions are ignored.

\begin{figure}
    \centering
\includegraphics[width=\columnwidth]{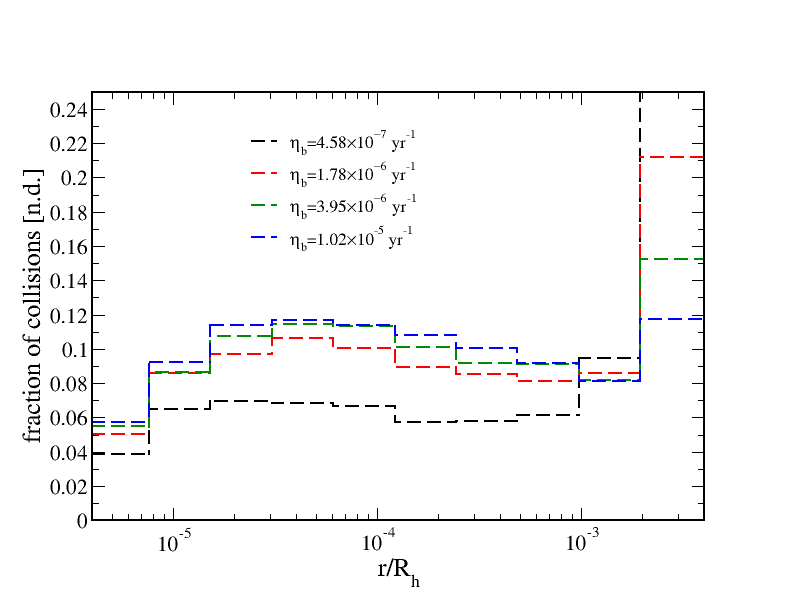}
    \caption{The fraction of collisions as a function of distance from the SMBH out of the total number of collisions, averaged over the later half of the simulations. Different curves correspond to the different injection rates shown in Figure \ref{fig:Allrates_bin}, as noted in the legend.}
    \label{fig:Colgrid_binnom}
\end{figure}

Another significant consequence of a high injection rate is on the distribution of distances from the SMBH where collisions occur. This is demonstrated in Figure \ref{fig:Colgrid_binnom}, which shows the resulting distributions (averaged over the second half of the simulation) for the different injection rates presented in Figure \ref{fig:TDE+DC+EMRIs_bin_colnom}.  

Higher injection rates naturally lead to a higher fraction of collisions which occur at smaller distances from the SMBH. Note that if at the injection radius the collision time scale is shorter than any evolutionary time scale (relaxation or GW emission), the distribution of collision distances essentially reflects the assumed distribution of the binaries' original separation distances. The log-normal distribution of $a_b$ between $a_{min}=0.01$au and $a_{max}=1$au leads to collisions of injected stars being spread roughly in a log-normal fashion in the range $10-1000\;$au from the SMBH. 

This result certainly reflects our simplifying assumption that disruption always occurs exactly so that $r_p=R_{Tb}(a_b)$. In reality there is a finite probability for disruption over a range of distances from the SMBH, which depends also on the distance of closest approach and orbit orientation \citep{Hills1988,Bromleyetal2006,Sarietal2010,Bromleyetal2012}. A full analysis of this effect is beyond the scope of this work, but we do expect our general results to be valid as long as the typical variability is much smaller than the total range of values for $a_b$. For example, \cite{Bromleyetal2006} demonstrated one case where for a given value of $a_b$, the resulting distribution of energies for the captured star is a Gaussian with a standard deviation of about $40\%$ ($20\%$ in the velocity of the ejected star). Such a variability, which is both symmetric and relatively small (compared to the range of binary separation distances), should not significantly affect the spatial distribution of collision radii, especially since stars are always captured in very eccentric orbits. In such orbits the collision time is shorter than the relaxation time (even for stars whose apoapse lies slightly above $R_{col}$), and  collisions mostly occur close to the periapse. It is noteworthy that allowing for a probabilistic outcome in binary disruption also implies a finite probability for tighter binaries not to be disrupted at all. Some of the tightest binaries will be destroyed as a combined TDE \citep{MandelLevin2015}, modifying the balance between TDEs and DCs.   

Our results suggest that collisions become dominant in the total collision rate already around  $\eta_b\approx 10^{-6}\invyr$. We emphasize that for $\eta_b\approx 10^{-5}\invyr$ collisions at and below $R\sim 100\;$au make up more than $60\%$ of the total collisions. This radius roughly corresponds to total kinetic energies of $10^{51}\;$ erg, comparable to supernovae in terms of the driving energy source. Our result concurs with the analytic estimate by \cite{BalbergSariLoeb2013}, and implies that for steady-state binary disruption rates of order $10^{-5}\invyr$, a Milky Way-like galaxy should exhibit "collisional supernovae" with similar rates as TDES. 

\begin{figure}
    \centering
\includegraphics[width=\columnwidth]{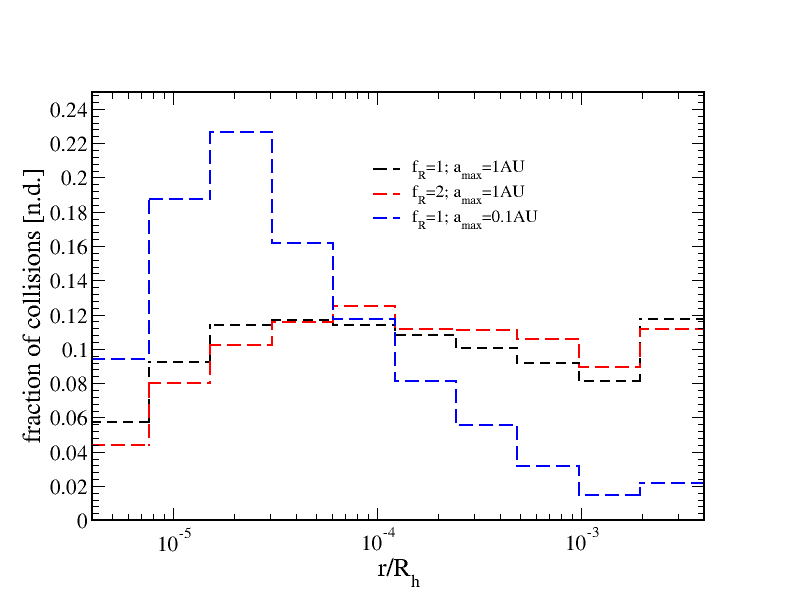}
    \caption{The fraction of collisions as a function of distance from the SMBH out of the total number of collisions, averaged over the later half of the simulations, for an injection rate of $\eta_b=10^{-5}\;\invyr$. Shown are the results for the nominal values of $f_R=1,\; a_{max}=1\;$au, and also for the variations $f_R=2$ or $f_R=1$ $a_{max}=0.1\;$au.}
    \label{fig:Colgrid_binfR2amx01AU}
\end{figure}

Applying the parameter survey ($f_R=2$ or $a_{max}=0.1\;$au) does indicate an important sensitivity in terms of the distribution of distances from the SMBH where DCs occur. Figure \ref{fig:Colgrid_binfR2amx01AU} shows the relative fraction of collisions which occur at different distances from the SMBH for simulations with the two variations, in both cases adjusted to yield a steady-state injection rate of $\eta_B=10^{-5}\invyr$.  

While the change to $f_R=2$ only slightly modifies the distribution of distances (again, it simply rescales the local densities), setting $a_{max}=0.1\;$au funnels the supply of injected stars to smaller distances from the SMBH, thus enhancing the DC rate at such radii. This implies that while the total DC rate is determined by $\eta_b$, the energy distribution function associated with collisions becomes highly localized. In particular, about one half of the collisions correspond to energies of order $10^{52}\;$erg. This result, if realistic, has significant implications concerning the prospects of observing the early optical flare that will follow a collision. Some caution is called for with regards to this conclusion, since if only the tightest binaries are candidates for generating injected stars, their fraction among the overall population in the cusp may be small, resulting in lower injection rates.   

\begin{figure}
    \centering
\includegraphics[width=\columnwidth]{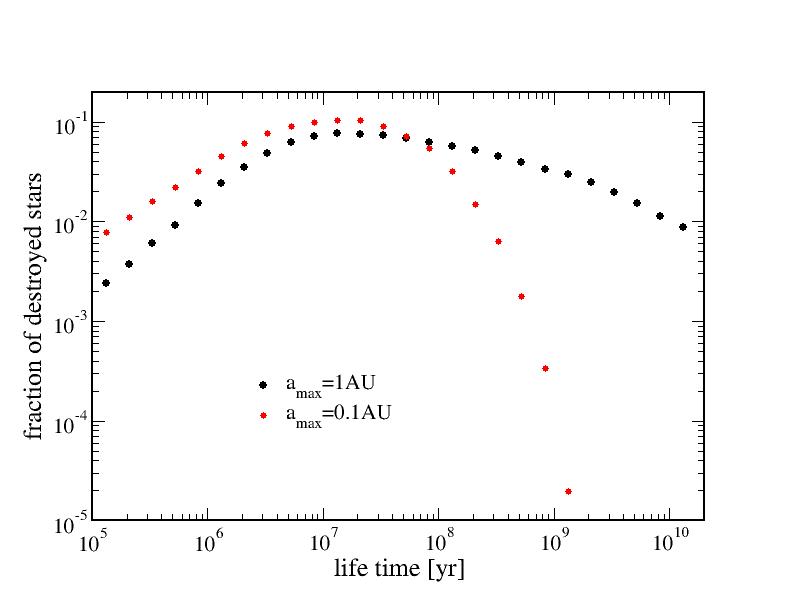}
    \caption{The fraction of lifetimes experienced by the injected stars, for simulations with collisions and $\eta_b=10^{-5}\;\invyr$. For this injection rate over $90\%$ of the injected stars are destroyed through collisions. Shown are the results for the nominal values of $f_R=1$, while varying the maximum initial binary separation distance by $a_{max}=1\;$au (black) $a_{max}=0.1\;$au (red).} 
    \label{fig:lifetime_binCOLandTITE}
\end{figure}

The role of the maximum binary separation distance is also reflected in the lifetime of injected stars. The distribution function of lifetimes for the two cases ($f_R=1,a_{max}=1.0\;$au) and ($f_R=1,a_{max}=0.1\;$au) over the second half of the simulation is shown in Figure \ref{fig:lifetime_binCOLandTITE}. Since more than $90\%$ of the injected stars are destroyed by collisions in both cases, we only show the total distribution function (as opposed to the breakdown of destruction channels shown in Figure \ref{fig:lifetime_binNOCOL}).

We find that in both cases the bulk of the lifetime distribution function centers around times of order $10^7-10^8\;$years. Since collisions deplete the density of stars in the inner part of the cusp, the two-body gravitational timescales the injected stars experience are longer than in a BW profile. The overabundance of shorter ($\sim10^5-10^6\;$yr) lifetimes, seen in Figure \ref{fig:lifetime_binNOCOL}, is therefore absent here, consistent with the suppression of the TDE and EMRI channels for injected stars. We note that setting  $a_{max}=0.1\;$au confines the injected stars to a tight range of radii, and so decreases the spread of lifetimes. Quantitatively we find that the average lifetime of an injected star is about $6\times 10^8\;$yr for $a_{max}=1\;$au, and reduces to about $10^8\;$yr for $a_{max}=0.1\;$au. Since these steady-state lifetimes correspond to an injection rate $\eta_b=10^{-5}\;\invyr$, these results imply that the momentary number of injected stars in the profile is about 6000 for $a_{max}=1\;$au, and 1000 for $a_{max}=0.1\;$au.

\begin{figure}
    \centering
\includegraphics[width=\columnwidth]{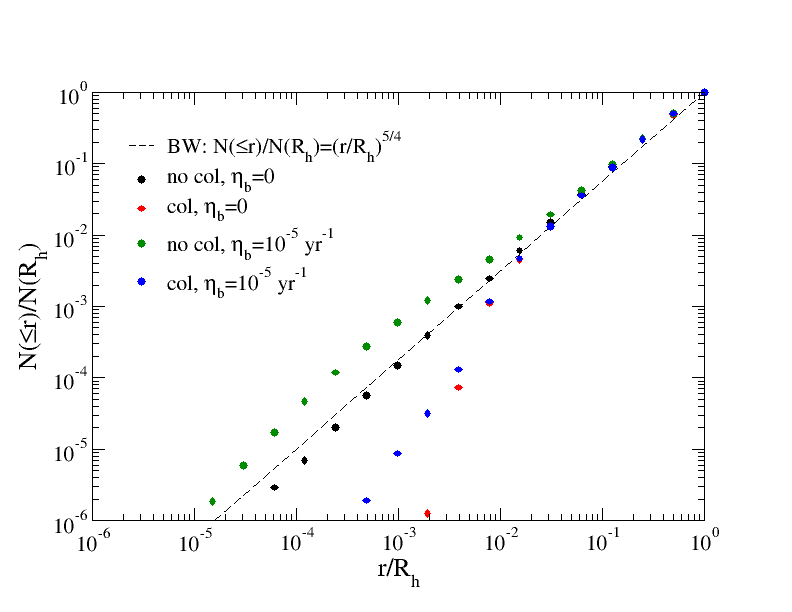}
    \caption{The cusp profile without ($\eta_b=0$) and with stars form disrupted binaries (injected at a rate $\eta_b=10^{-5}\;\invyr$), without collisions (black and green symbols, respectively) and again with collisions (red and blue symbols, respectively). Shown is $N(\leq r)$, the total number of stars with sma $\leq r_i$, averaged over the last $10\%$ of the simulations. Also shown (dashed line) is the theoretical profile of \cite{BW76}, $N(\leq r)=N\times(r/R_h)^{5/4}$, which does not account for disrupted binaries, collisions or relativistic effects on the orbits. Stars are injected with $a_{max}=1\;$au, and collisions are estimated with $f_R=1$ in the collision cross section.}
    \label{fig:Profiles}
\end{figure}

To complete the analysis, we present in Figure \ref{fig:Profiles} the steady-state profile of the cusp in terms of $N(r)$ for four representative cases. This profile is calculated by averaging the vector $N(i)$ over the last $10\%$ of the simulation time. The four cases are (i) no disrupted binaries and no collisions, (ii) stars injected due to disrupted binaries with $\eta_b=10^{-5}\;\invyr$ but no collisions, (iii) no disrupted binaries and including collisions, and (iv) injected stars at $\eta_b=10^{-5}\;\invyr$ and including collisions.
We see that without disrupted binaries and without collisions the profile tends to maintain the original BW $N(\leq r)\propto r^{5/4}$ result. The innermost part of the cusp is slightly underpopulated, as is to be expected due to the dilution caused by the accelerated losses through GW emission. When the injection rate is significant (but still ignoring DCs), there is an obvious excess at small radii with respect to the BW profile, since stars are injected specifically into the inner cusp, at the expense of the region near $R_h$. These results are naturally consistent with the original analysis by \cite{SariFragione2019}. When collisions are taken into account the interior part of the cusp settles into a different steady state, while the exterior (near $R_h$) steady-state profile is unchanged (and depends only on the mechanisms which supply stars to the outer cusp). Collisions clearly create a significant depletion mechanism for the inner part of the cusp. Without injected stars, the main source of stars for this region is through the diffusion of stars from larger radii, so the steady state is achieved through collisions close to the (fiducial) value of $R_{col}\approx 8\;$mpc set in the simulation. 
Since most stars arrive at $r\sim R_{col}$ along quasi-circular orbits, $N(r)$ drops over a very short range from the BW slope to practically zero (see Section \ref{subsec:collisions}). When the injection rate is large, it establishes a steady population which balances DCs in the inner cusp and $N(r)$ there is significant. Moreover, recall that injected stars have highly eccentric orbits, so most collisions occur closer to the periapse, rather than at the sma (compare the distribution of collision distances in Figure \ref{fig:Colgrid_binfR2amx01AU} to the $N(r)$ profile).

\section{Summary and Conclusions}\label{sec:Summary}

Recent and upcoming advances in time-domain astronomy offer new opportunities for studying energetic transient sources. Dedicated surveys, such as ZTF \footnote{https://www.ztf.caltech.edu/} \citep{zTF} and ASAS-SN \footnote{https://www.astronomy.ohio-state.edu/asassn/} \citep{ASASSN} have dramatically increased detection rates for energetic high-cadence events, and these will be enhanced further in terms of discovery rates and observed wavelength (near UV) by the upcoming ULTRASAT \footnote{https://www.weizmann.ac.il/ultrasat/} mission \citep{ULTRASAT}. The prospects of detecting luminous and rare transients in general are on the rise, and these include sources which are unique to galactic nuclei where an SMBH reigns over the dynamics and the fate of the stars surrounding it.   

In galactic nuclei harboring an SMBH, the stellar orbits are dominated by the deep potential well of the SMBH, while two-body interactions among the stars modify their energy and angular momentum, and shape their distribution over longer timescales. Stars may be scattered into extremely eccentric orbits, which will occasionally result in a star being tidally disrupted if it ventures too close to the SMBH. Moreover, stars sufficiently close to the SMBH also evolve their orbits through energy loss due to GW radiation emission. If this energy loss is rapid enough, these stars will inspiral individually toward the SMBH on a general relativistic timescale, and disrupt "gradually" as they approach and cross the tidal disruption radius. In this case, a possible electromagnetic flare will be accompanied with a GW event with emission in the millihertz frequency band \citep{Alexander2017}.

In this work, we examined the rates of high-velocity destructive stellar collisions which occur close to the SMBH. Such collisions compete with the other destructive channels. We used a simplified numerical Monte Carlo analysis of stellar dynamics in galactic stellar nuclei. We applied the specific version by \cite{SariFragione2019} for simulating two-body relaxation and GW emission effects on stars orbiting an SMBH, to which we added self-consistent tracking of DCs. Our main results concern the relative rates of the different channels of stellar destruction near the SMBH, which are TDEs, EMRIs, and DCs. 

We find that for a gravitationally relaxed Milky Way-like stellar cluster and SMBH, the DC rate is of order a few percent of the TDE rate. The latter is dominated by the scattering of stars into eccentric orbits in the vicinity of the radius of influence, and the former is the main destruction mechanism of stars which diffuse into the inner part of the cusp. When compared to previous works which did not account for collisions, we find that DCs replace EMRIs as the secondary destruction channel in the cusp. Once collisions are included, main-sequence-star EMRIs become practically nonexistent (of course EMRIs of compact objects should not be affected).

We also examined the role of binary tidal disruptions in the destructive processes in galactic nuclei. For stars injected on highly eccentric orbits close to the SMBH due to Hills binary disruption, we find that they are almost exclusively destroyed through high-velocity collisions. In other words, the DC rate essentially settles on a steady state with the injection rate. The TDE rate is unaffected (since it is dominated by stars scattered into eccentric orbits far from the SMBH), and the EMRI rate remains very small (although binary disruption does increase this rate due to the finite probability of injected stars to avoid collisions and eventually inspiral toward the SMBH). For a Milky Way-like system with Sun-like stars, we find that injection rates of order $10^{5}\;\invyr$ will generate a DC rate which is comparable to the TDE rate. 

Again, it is noteworthy that these results significantly modify the expected EMRI rates for high injection rates when collisions are not considered. In such simulations (see also \cite{SariFragione2019}), injected stars have sizable probabilities to either be disrupted by the SMBH on plunging orbits or on gradually circularized orbits due to GW emission, so high injection rates predict comparable EMRI and TDE rates. Allowing for collisions is therefore crucial when considering the prospects of detecting luminous GW emission due to main-sequence-star EMRIs, or for interacting EMRIs to create quasiperiodic emissions \citep{Metzgeretal2022}. 

An important aspect of our results is the spatial distribution of high-velocity stellar collisions. In our simplified model we assumed that collisions are "destructive" when the typical velocity, estimated according to the SMBH potential, is greater than the stellar escape velocity. For Sun-like stars DCs correspond to energies of order $\sim 10^{49}\;$erg and higher. In terms of the distance from the SMBH, for Milky-Way parameters this corresponds to a distance of order $R_{col}\leq 8\;$mpc. In a steady state, gravitationally relaxed system, stars are supplied to the inner part of the cusp through diffusion in energy and angular momentum space; in other words, from the outer cusp to its interior. As a result, almost all collisions occur close to $R_{col}$, which implies that the optical flare following the collision will be very dim, and the debris will, at best, accrete very slowly onto the SMBH. The observational potential of such collisions is poor. However, stars from disrupted binaries are injected much closer to the SMBH than $R_{col}$, and their collisions will respectively be more energetic. We reproduce the rough analytic estimate of \cite{BalbergSariLoeb2013}, that most DCs of injected stars will be at energies of order $10^{51}-10^{52}\;$erg, implying a short but bright optical flare. These are "collisional supernovae", whose rate might be comparable to the TDE rate, if the binary disruption rate is significant.

Some simplifications in our model should be highlighted as points for further consideration. The Monte Carlo two-body averaged scheme cannot take into account coherent torques between slowly precessing orbits, namely resonant relaxation processes \citep{KocsisTremaine2011,KocsisTremaine2015}. On average resonant relaxation is relevant at radii beyond the DC regime ($r\gtrsim 1000$ au), since at smaller distances relativistic precession decouples the GW inspiral from the residual torques of the background stars. This result should be reevaluated when considering a stellar profile which is modified by injected stars. We plan to examine this issue separately with designated $N$-body simulations (G.~Yassur and S.~Balberg, 2023, in preparation).

In our model we also ignored the role of low-velocity collisions in the outer parts of the stellar cusp. These are generally not destructive, and more likely result in mass transfer, mass loss, and mergers \citep{FreitagBenz2005}. Multiple collisions may be however be destructive, and in any case, changes in the number and masses of single stars may accumulate to quantitative effects on the dynamics and steady-state profile in the outer cusp \citep{Sillsetal2005,DaleDavies2006,Roseetal2023}. In particular, they might enhance the formation of stellar mass black holes \citep{Roseetal2022}, which will eventually affect the cusp dynamics.

Finally, we clearly oversimplified our analysis by considering a population of single-mass stars. Realistically, there should be a mass function both for the stars in the cusp and the injected stars, which implies a complex relation between the mass and number densities. Moreover, a full mass function will result in segregation due to dynamical friction, and objects with larger masses will generally sink toward the center thus adjusting the cusp density profile. Most notably, there is likely to be segregation of compact objects towards the inner part of the cusp \citep{KHA2009,LinialSari2022}, with stellar black holes leading with the steepest density profile. Compact objects will affect main-sequence stars dynamically by shortening the two-body relaxation timescales, without participating in collisions  (although a compact object may still disrupt a main-sequence star, depending on the impact velocity; see the appendix in \cite{MetzgerStone2017}). Thus a large population of compact remnants in the inner cusp may change the relative mix of DCs and main-sequence EMRIs. The dynamics and destruction rates can be complicated even further if one or more intermediate-mass ($\sim 10^{3}M_\odot$) black hole is present in the inner cusp  \citep{fgk18,flgk18}. We address some these complications with $N$-body simulations of the inner cusp in our separate work (G.~Yassur and S.~Balberg, 2023, in preparation).

\section{Acknowledgements}

We thank Reem Sari and Barak Rom for useful comments and discussions.

\bibliography{Collisionsbib}{}

\begin{thebibliography}{}
\expandafter\ifx\csname natexlab\endcsname\relax\def\natexlab#1{#1}\fi
\providecommand{\url}[1]{\href{#1}{#1}}
\providecommand{\dodoi}[1]{doi:~\href{http://doi.org/#1}{\nolinkurl{#1}}}
\providecommand{\doeprint}[1]{\href{http://ascl.net/#1}{\nolinkurl{http://ascl.net/#1}}}
\providecommand{\doarXiv}[1]{\href{https://arxiv.org/abs/#1}{\nolinkurl{https://arxiv.org/abs/#1}}}

\bibitem[{{Alexander}(2017)}]{Alexander2017}
{Alexander}, T. 2017, \araa, 55, 17,
  \dodoi{10.1146/annurev-astro-091916-055306}

\bibitem[{{Amaro-Seoane}(2023)}]{AmaroSeoane2023}
{Amaro-Seoane}, P. 2023, \apj, 947, 8, \dodoi{10.3847/1538-4357/acb8b9}

\bibitem[{{Arnold} {et~al.}(2022){Arnold}, {Baumgardt}, \& {Wang}}]{Nbody6pp}
{Arnold}, A.~D., {Baumgardt}, H., \& {Wang}. 2022, \mnras, 509, 2075,
  \dodoi{10.1093/mnras/stab3090}

\bibitem[{{Bahcall} \& {Wolf}(1976)}]{BW76}
{Bahcall}, J.~N., \& {Wolf}, R.~A. 1976, \apj, 209, 214, \dodoi{10.1086/154711}

\bibitem[{{Bahcall} \& {Wolf}(1977)}]{BW77}
---. 1977, \apj, 316, 883, \dodoi{10.1086/155534}

\bibitem[{{Balberg} {et~al.}(2013){Balberg}, {Sari}, \&
  {Loeb}}]{BalbergSariLoeb2013}
{Balberg}, S., {Sari}, R., \& {Loeb}, A. 2013, \mnras, 434, L26,
  \dodoi{10.1093/mnrasl/slt071}

\bibitem[{{Bar-Or} \& {Alexander}(2016)}]{BarOrAlexander2016}
{Bar-Or}, B., \& {Alexander}, T. 2016, \apj, 820, 129,
  \dodoi{10.3847/0004-637X/820/2/129}

\bibitem[{{Bellm} {et~al.}(2019){Bellm}, {Kulkarni}, \& {Graham}}]{zTF}
{Bellm}, E.~C., {Kulkarni}, S.~R., \& {Graham}, M.~J. e.~a. 2019, \pasp, 131,
  018002, \dodoi{10.1088/1538-3873/aaecbe}

\bibitem[{{Binney} \& {Tremaine}(1987)}]{BinneyTremaine1987}
{Binney}, J., \& {Tremaine}, S. 1987, {Galactic Dynamics} (Princeton NJ:
  Princeton University Press), \dodoi{1987gady.book.....B}

\bibitem[{{Bromley} {et~al.}(2006){Bromley}, {Kenyon}, {Geller}, {Barcikowski},
  {Brown}, \& {Kurtz}}]{Bromleyetal2006}
{Bromley}, B.~C., {Kenyon}, S.~J., {Geller}, M.~J., {et~al.} 2006, \apj, 653,
  1194, \dodoi{10.1086/508419}

\bibitem[{{Bromley} {et~al.}(2012){Bromley}, {Kenyon}, {Geller}, \&
  {Brwon}}]{Bromleyetal2012}
{Bromley}, B.~C., {Kenyon}, S.~J., {Geller}, M.~J., \& {Brwon}, W.~R. 2012,
  \apjl, 749, 42, \dodoi{10.1088/2041-8205/749/2/L42}

\bibitem[{{Brown}(2015)}]{Brown2015}
{Brown}, Warren, R. 2015, \araa, 53, 15,
  \dodoi{10.1146/annurev-astro-082214-122230}

\bibitem[{{Dale} \& {Davies}(2006)}]{DaleDavies2006}
{Dale}, J.~E., \& {Davies}, M.~B. 2006, \mnras, 366, 1424,
  \dodoi{10.1111/j.1365-2966.2005.09937.x}

\bibitem[{{Duchêne} \& {Kraus}(2013)}]{DucheneKraus2013}
{Duchêne}, G., \& {Kraus}, A. 2013, \araa, 51, 269,
  \dodoi{10.1146/annurev-astro-081710-102602}

\bibitem[{{Fabian}(2012)}]{AGNII}
{Fabian}, A.~R. 2012, \araa, 50, 455,
  \dodoi{10.1146/annurev-astro-081811-125521}

\bibitem[{{Fouvry} {et~al.}(2022){Fouvry}, {Dehnen}, {Tremaine}, \&
  {Bar-Or}}]{Fouvryetal2022}
{Fouvry}, J.-B., {Dehnen}, W., {Tremaine}, S., \& {Bar-Or}, B. 2022, \apj, 931,
  8, \dodoi{10.3847/1538-4357/ac602e}

\bibitem[{{Fragione} \& {Antonini}(2019)}]{FragioneAntonini2018}
{Fragione}, G., \& {Antonini}, F. 2019, \mnras, 488, 728,
  \dodoi{10.1093/mnras/stz1723}

\bibitem[{{Fragione} \& {Sari}(2018)}]{FragioneSari2018}
{Fragione}, G., \& {Sari}, R. 2018, \apj, 852, 51,
  \dodoi{10.3847/1538-4357/aaa0d7}

\bibitem[{{Freitag} \& {Benz}(2005)}]{FreitagBenz2005}
{Freitag}, M., \& {Benz}, W. 2005, \mnras, 358, 1133,
  \dodoi{10.1111/j.1365-2966.2005.08770.x}

\bibitem[{{Frgaione} {et~al.}(2018{\natexlab{a}}){Frgaione}, {Ginzburg}, \&
  Bence}]{fgk18}
{Frgaione}, G., {Ginzburg}, I., \& Bence, K. 2018{\natexlab{a}}, \apj, 856, 92,
  \dodoi{10.3847/1538-4357/aab368}

\bibitem[{{Frgaione} {et~al.}(2018{\natexlab{b}}){Frgaione}, {Leigh}, \&
  Bence}]{flgk18}
{Frgaione}, G., {Leigh}, Nathan~W.~C.{Ginzburg}, I., \& Bence, K.
  2018{\natexlab{b}}, \apj, 867, 119, \dodoi{10.3847/1538-4357/aae486}

\bibitem[{{Gezari}(2021)}]{Gezari2021}
{Gezari}, S. 2021, \araa, 59, 21, \dodoi{10.1146/annurev-astro-111720-030029}

\bibitem[{{Heckmn} \& {Best}(2014)}]{HeckmanBest2014}
{Heckmn}, T.~M., \& {Best}, P.~N. 2014, \araa, 52, 589,
  \dodoi{10.1146/annurev-astro-081913-035722}

\bibitem[{{Heggie} \& {Hut}(1993)}]{HeggieHut1993}
{Heggie}, D.~C., \& {Hut}, P. 1993, \apjs, 85, 347, \dodoi{10.1086/191768}

\bibitem[{{Henon}(1971)}]{Henon1971}
{Henon}, M, H. 1971, \apss, 14, 151, \dodoi{10.1007/BF00649201}

\bibitem[{{Hills}(1988)}]{Hills1988}
{Hills}, J.~C. 1988, \nat, 331, 687, \dodoi{10.1038/331687a0}

\bibitem[{{Hopman}(2009)}]{Hopman2009}
{Hopman}, C. 2009, \apj, 700, 1933, \dodoi{10.1088/0004-637X/700/2/1933}

\bibitem[{{Hopman} \& {Alexander}(2005)}]{HopmanAlexander2005}
{Hopman}, C., \& {Alexander}, T. 2005, \apj, 629, 362, \dodoi{10.1086/431475}

\bibitem[{{Hopman} \& {Alexander}(2006)}]{HopmanAlexander2006}
---. 2006, \apj, 645, 1152, \dodoi{10.1086/504400}

\bibitem[{{Keshet} {et~al.}(2009){Keshet}, {Hopman}, \& {Alexander}}]{KHA2009}
{Keshet}, U., {Hopman}, C., \& {Alexander}, T. 2009, \apjl, 698, L64,
  \dodoi{10.1088/0004-637X/698/1/L64}

\bibitem[{{Kochanek} {et~al.}(2017){Kochanek}, {Shappee}, \& {Satnek}}]{ASASSN}
{Kochanek}, C.~S., {Shappee}, B.~J., \& {Satnek}, K.~Z. e.~a. 2017, \pasp, 129,
  104502, \dodoi{10.1088/1538-3873/aa80d9}

\bibitem[{{Kocsis} \& {Tremaine}(2011)}]{KocsisTremaine2011}
{Kocsis}, B., \& {Tremaine}, S. 2011, \mnras, 412, 187,
  \dodoi{10.1111/j.1365-2966.2010.17897.x}

\bibitem[{{Kocsis} \& {Tremaine}(2015)}]{KocsisTremaine2015}
---. 2015, \mnras, 448, 3265, \dodoi{10.1093/mnras/stv057}

\bibitem[{{Kormendy} \& {Ho}(2013)}]{KormendyHo2013}
{Kormendy}, J., \& {Ho}, L.~c. 2013, \araa, 51, 511,
  \dodoi{10.1146/annurev-astro-082708-101811}

\bibitem[{{Lightman} \& {Shapiro}(1977)}]{LightmanShapiro1977}
{Lightman}, A.~P., \& {Shapiro}, S.~L. 1977, \apj, 211, 244L,
  \dodoi{10.1086/154925}

\bibitem[{{Linial} \& {Sari}(2022)}]{LinialSari2022}
{Linial}, I., \& {Sari}, R. 2022, \apjl, 940, L101,
  \dodoi{10.3847/1538-4357/ac9bfd}

\bibitem[{{Linial} \& {Sari}(2023)}]{LinialSari2023}
---. 2023, \apjl, 945, L86, \dodoi{10.3847/1538-4357/acbd3d}

\bibitem[{{Mandel} \& {Levin}(2015)}]{MandelLevin2015}
{Mandel}, I., \& {Levin}, Y. 2015, \apjl, 805, L4,
  \dodoi{10.1088/2041-8205/805/1/L4}

\bibitem[{{Merritt}(2013)}]{Merritt2013}
{Merritt}, D. 2013, {Dynamics and Evolution of Galactic Nuclei} (Princeton NJ:
  Princeton University Press), \dodoi{2013degn.book.....M}

\bibitem[{{Merritt} \& {Ferrarese}(2001)}]{MerrittFerrarese2001}
{Merritt}, D., \& {Ferrarese}, L. 2001, \apj, 547, 140, \dodoi{10.1086/318372}

\bibitem[{{Metzger} \& {Stone}(2017)}]{MetzgerStone2017}
{Metzger}, B.~D., \& {Stone}, N.~C. 2017, \apj, 844, 75,
  \dodoi{10.3847/1538-4357/aa7a16}

\bibitem[{{Metzger} {et~al.}(2022){Metzger}, {Stone}, \&
  {Gilbaum}}]{Metzgeretal2022}
{Metzger}, B.~D., {Stone}, N.~C., \& {Gilbaum}, S. 2022, \apj, 926, 101,
  \dodoi{10.3847/1538-4357/ac3ee1}

\bibitem[{{Perets} \& {\v{S}ubr}(2012)}]{PeretsSubr2012}
{Perets}, H.~B., \& {\v{S}ubr}, L. 2012, \apj, 751, 133,
  \dodoi{10.1088/0004-637X/751/2/133}

\bibitem[{{Peters}(1964)}]{Peters1964}
{Peters}, P.~C. 1964, Physical Review, 136, 1224,
  \dodoi{10.1103/PhysRev.136.B1224}

\bibitem[{{Rauch} \& {Tremaine}(1996)}]{RauchTremaine1996}
{Rauch}, K.~P., \& {Tremaine}, S. 1996, New Astronomy, 1, 149,
  \dodoi{10.1016/S1384-1076(96)00012-7}

\bibitem[{{Rees}(1984)}]{AGNI}
{Rees}, M.~J. 1984, \araa, 22, 481, \dodoi{10.1146/annurev.aa.22.090184.002351}

\bibitem[{{Rees}(1988)}]{Rees1988}
---. 1988, \nat, 333, 523, \dodoi{10.1038/333523a0}

\bibitem[{{Rose} {et~al.}(2022){Rose}, {Naoz}, {Sari}, \&
  {Linial}}]{Roseetal2022}
{Rose}, S.~C., {Naoz}, S., {Sari}, R., \& {Linial}, I. 2022, \apjl, 929, L22,
  \dodoi{10.3847/2041-8213/ac6426}

\bibitem[{{Rose} {et~al.}(2023){Rose}, {Naoz}, {Sari}, \&
  {Linial}}]{Roseetal2023}
---. 2023, ApJ., submitted.
\newblock \doarXiv{2304.10569}

\bibitem[{{Rossi} {et~al.}(2014){Rossi}, {Kobayashi}, \&
  {Sari}}]{Rossietal2014}
{Rossi}, E.~M., {Kobayashi}, S., \& {Sari}, R. 2014, \apj, 795, 125,
  \dodoi{10.1088/0004-637X/795/2/125}

\bibitem[{{Sari} \& {Fargione}(2019)}]{SariFragione2019}
{Sari}, R., \& {Fargione}, G. 2019, \apj, 885, 1,
  \dodoi{10.3847/1538-4357/ab43df}

\bibitem[{{Sari} \& {Rossi}(2010)}]{Sarietal2010}
{Sari}, Reem~{Kobayashi}, S., \& {Rossi}, E.~M. 2010, \apj, 708, 605,
  \dodoi{10.1088/0004-637X/708/1/605}

\bibitem[{{Shvartsvald} {et~al.}(2023){Shvartsvald}, {Waxman}, \&
  {Gal-Yam}}]{ULTRASAT}
{Shvartsvald}, Y., {Waxman}, E., \& {Gal-Yam}, Avishai, e.~a. 2023, Submitted
  to the AAS journals, \dodoi{https://arxiv.org/abs/2304.14482}

\bibitem[{{Sills} {et~al.}(2005){Sills}, {Adams}, \& {Davies}}]{Sillsetal2005}
{Sills}, A.~E., {Adams}, T., \& {Davies}, M.~B. 2005, \mnras, 358, 716,
  \dodoi{10.1111/j.1365-2966.2005.08809.x}

\end{thebibliography}
\bibliographystyle{aasjournal}

\end{document}